\documentclass{article}
\usepackage{arxiv}
\usepackage[utf8]{inputenc} 
\usepackage[T1]{fontenc}    
\usepackage{hyperref}       
\usepackage{url}            
\usepackage{booktabs}       
\usepackage{amsfonts}       
\usepackage{nicefrac}       
\usepackage{microtype}      
\usepackage{graphicx}
\usepackage{doi}
\usepackage{amsmath,amssymb,amsfonts}
\usepackage{algorithm}
\usepackage{algpseudocode}
\usepackage{textcomp}
\usepackage{caption}
\usepackage{subcaption}
\usepackage{wrapfig}
\usepackage{color}
\usepackage{vcell}
\usepackage{multicol}
\usepackage{multirow}

\title{Medical Image Data Provenance for Medical Cyber-Physical System}

\author{
    \href{https://orcid.org/0000-0001-6840-7913}{\includegraphics[scale=0.06]{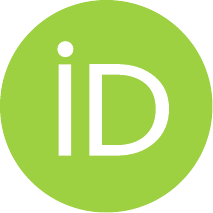}\hspace{1mm}Vijay Kumar}\\
    ANSK School of Information Technology\\
    Indian Institute of Technology Delhi\\
    India, 110016\\
    \texttt{vijay.kumar@sit.iitd.ac.in}
    \And
    \href{https://orcid.org/0000-0000-0000-0000}{\includegraphics[scale=0.06]{orcid.pdf}\hspace{1mm}Kolin Paul} \\
    Department of Computer Science and Engineering\\
    Indian Institute of Technology Delhi\\
    India, 110016\\
    \texttt{kolin@cse.iitd.ac.in} \\
}

\renewcommand{\shorttitle}{Medical Image Data \textit{Provenance} for Medical Cyber-Physical Systems}

\hypersetup{
pdftitle={Medical Image Data Provenance for Medical Cyber-Physical System},
pdfsubject={cs.CV, cs.CR, cs.ET, eess.IV, eess.SP},
pdfauthor={V.~Kumar, K.~Paul},
pdfkeywords={Telemedicine, Image Watermarking, Data Provenance, Device Fingerprint, Medical Internet of Things (IoMT), mHealth/eHealth, Data integrity},
}

\begin{document}
\maketitle

\begin{abstract}
Continuous advancements in medical technology have led to the creation of affordable mobile imaging devices suitable for telemedicine and remote monitoring. However, the rapid examination of large populations poses challenges, including the risk of fraudulent practices by healthcare professionals and social workers exchanging unverified images via mobile applications. To mitigate these risks, this study proposes using watermarking techniques to embed a device fingerprint (DFP) into captured images, ensuring data provenance. The DFP, representing the unique attributes of the capturing device and raw image, is embedded into raw images before storage, thus enabling verification of image authenticity and source. Moreover, a robust remote validation method is introduced to authenticate images, enhancing the integrity of medical image data in interconnected healthcare systems. Through a case study on mobile fundus imaging, the effectiveness of the proposed framework is evaluated in terms of computational efficiency, image quality, security, and trustworthiness. This approach is suitable for  a range of applications including telemedicine, the Internet of Medical Things (IoMT), eHealth and Medical Cyber-Physical Systems (MCPS) applications, providing a reliable means to maintain data provenance in diagnostic settings utilizing medical images or videos.
\end{abstract}


\section{Introduction}\label{sec:Background and Related Work}
Data provenance in healthcare refers to documenting and tracking the origin, transformations, and movement within the healthcare system~\cite{glavic2021data,johns2023data,martinez2022comprehensive,pan2023data}. It ensures transparency, traceability, and reliability in healthcare data, which is critical for maintaining the integrity and trustworthiness of information in an interconnected healthcare system. Data provenance enhances accountability, facilitates error detection, and supports regulatory compliance in healthcare systems by recording the history of data creation, modification, and transmission. For data provenance, many frameworks are currently available that use four primary techniques commonly, namely (1) logging-based technology, (2) cryptography-based technology, (3) blockchain-based technology, and (4) ontology-based technology~\cite{johns2023data,pan2023data,hu2020survey,9842712,ahmed2023data}. These techniques are often employed to achieve data provenance by carrying origin information, source information, and traceability.
\par
In recent years, medical imaging has become more crucial, allowing important health information to be shared with remote devices, servers, or specialists in the healthcare system~\cite{martinez2022comprehensive,vijay2023fundus,mittal2020computerized,newaz2021survey}. This information includes medical prescriptions, pathological reports, images and videos collected by imaging-based diagnosis systems, and postures and activities monitored via images and video, etc.~\cite{gull2023advances,pan2009medical}. As a result, healthcare professionals, researchers, and analysts use images to improve patient care and medical knowledge. However, the sensitive nature of these patients' images, ownership, and ethical use of healthcare image data is crucial. Like other healthcare data, patient images are always at risk of security breaches. Therefore, it is crucial to establish a comprehensive framework for data provenance to ensure the safety and reliability of healthcare image data.
\par 
As a result, various techniques have been proposed, including digital signatures, watermarking, metadata, certificates, time-stamping, hashing, blockchain, and machine learning~\cite{johns2023data,ahmed2023data,kohli2023intelligent,nabi2022comprehensive}. However, each technique has its limitations. Digital signatures are vulnerable to key compromise attacks and require secure key management~\cite{martinez2022comprehensive,pan2009medical,kohli2023intelligent}. Watermarking can degrade image quality, and attackers can easily alter or remove metadata~\cite{pan2009medical,wan2022comprehensive}. Certificates verify sender identity but have limitations in detecting tampering~\cite{martinez2022comprehensive,newaz2021survey}. Time-stamping can't detect tampering after timestamp creation, and hashing is vulnerable to collision attacks. Blockchain is promising but faces challenges in scalability and interoperability~\cite{mcbee2020blockchain}. Machine learning requires significant training data and may not be effective against sophisticated attacks~\cite{rosenblatt2023enhancement}. Combinations of techniques are necessary for comprehensive provenance. Further research is needed to address these limitations and improve the effectiveness of these techniques.
\begin{figure}[t!]
    \centering
    \includegraphics[width=0.75\linewidth]{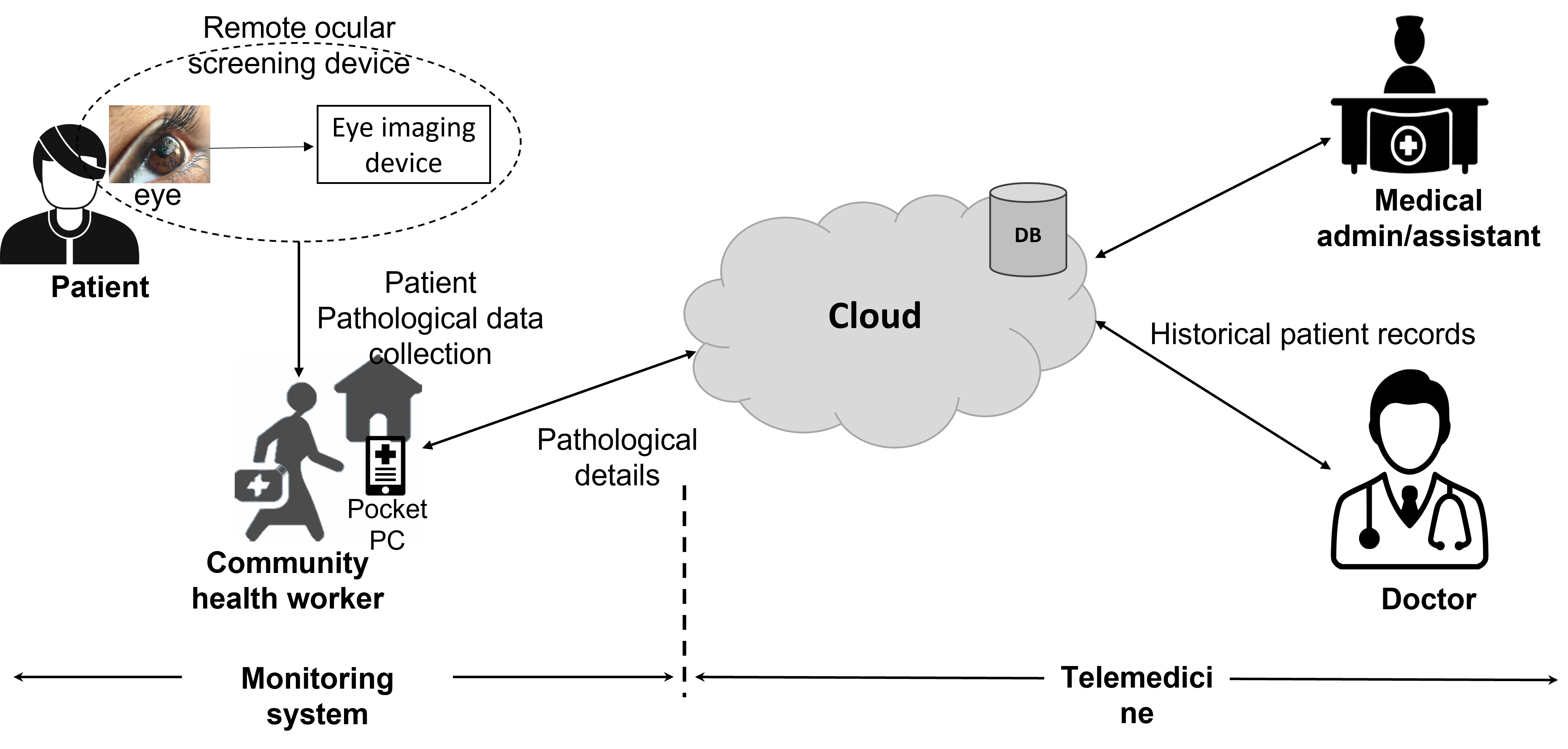}
    \caption{Retinal imaging-based affordable healthcare application for IoMT/mHealth.}
    \label{fig:mHealth application example of fundus imaging}
\end{figure}

\subsection{Medical Imaging and Data Provenance}
Telemedicine, a component of Medical Cyber-Physical Systems (MCPS) and the Internet of Medical Things (IoMT), revolutionizes healthcare delivery through remote clinical services enabled by information and communication technologies (ICT). Its impact is particularly pronounced in underserved areas with limited access to traditional healthcare resources. In the context of retinal and ocular diseases (shown in Figure~\ref{fig:mHealth application example of fundus imaging}), telemedicine ensures timely and accessible healthcare interventions~\cite{dasgupta2008telemedicine,kumar2016mnetra,lekha2019mii,paul2015fundus,singh2019innovative,toslak2018wide}. 
Additionally, with the popularity of smartphones and internet-enabled handheld devices, acquiring and sharing retinal images and videos has become more feasible. However, this task is particularly challenging in developing nations, where healthcare availability is limited, and the patient-to-doctor ratio is low~\cite{antaki2020role}. In such scenarios, social workers or healthcare agencies are often employed, but fraudulent activities can occur when healthcare and social workers exchange photographs obtained through unauthorized software or hardware. The impact of such fraud can be significant, affecting people's lives~\cite{glavic2021data,nabi2022comprehensive,KumbhMe2021fakeCovid,TransparTrust2020PublicComm,CanIndia2021covidDataTrust,fake2023MedicalReport}. 
\par   
In the past few years, many methods and frameworks for tracing the source of medical data have been developed. Embedding metadata information into images or videos is one of the most common practices~\cite{nabi2022comprehensive}. Metadata is usually included in all image and video files. It gives basic information about the image, like its resolution, imaging sensor, camera focal length, quantization level, and color space. In this instance, however, meta-data makes it simple for attackers or malicious users to quickly get sensitive information from images. Further, they used this information to alter the embedded information easily. This creates a significant risk to patient safety and medical treatment procedures. The COVID19 pandemic has highlighted the importance of detecting fake data injected by healthcare workers using mobile devices and medical applications. Maintaining data integrity and security is critical in the ocular or retinal healthcare domain, particularly for retinal imaging datasets. Fundus imaging datasets are used to diagnose and monitor various ocular diseases, and the metadata accompanying these images contains essential information about the imaging devices used. However, retinal imaging datasets are vulnerable to tampering and exchange by unauthorized users, posing a significant risk to patient safety and the accuracy of medical procedures~\cite{rosenblatt2023enhancement}. In addition to image or data tampering, the problem of device tampering or the use of counterfeit and compromised devices presents another significant challenge. Therefore, timely detection of such malicious activities within networks is crucial to prevent adverse effects.

\subsection{Consequences of Data Provenance Constraints in Healthcare}
The limitations and constraints of the different medical image data provenance techniques have several consequences on medical imaging-based affordable healthcare~\cite{johns2023data,martinez2022comprehensive,ahmed2023data,kohli2023intelligent}. Firstly, these limitations increase the risk of unauthorized access, alteration, or manipulation of medical images, which can seriously affect patient health outcomes. Secondly, the limitations of these techniques also hinder the ability of healthcare providers to detect and prevent fraud, which can lead to financial losses and compromise the integrity of medical research. Additionally, compromised datasets affect the quality of ongoing healthcare research, undermining research integrity. Consequently, many countries, including the USA (Health Insurance Portability and Accountability Act, HIPAA), the European Union (General Data Protection Regulation, GDPR), and India (Personal Health Records, PHR Standards, Information Technology Act, IT Act, and National Digital Health Mission-NDHM), emphasize the need for robust data provenance models to ensure the privacy and security of digital and electronic healthcare data~\cite{ahmed2023data,flaumenhaft2018personal,gudi2021challenges,roehrs2017personal}. Therefore, it is crucial to have a framework that uses advanced image processing and computer vision techniques, with a unique identification and tracing method, to identify fraudulent medical images shared by remote users~\cite{glavic2021data,pan2023data,ahmed2023data}. The framework aims to improve the accuracy and reliability of medical imaging in remote healthcare applications, enhance patient outcomes, and reduce the impact of fraudulent activities.

\subsection{Security Challenges in Medical Cyber-Physical Systems}
MCPS integrate medical devices, sensors, software, and networking technologies to monitor, diagnose, and treat patients. With advancements in technologies like the IoMT and eHealth, CPS-compatible medical equipment is becoming more prevalent, but it also introduces new security vulnerabilities ~\cite{alaa2019automated}. These vulnerabilities, such as device counterfeiting or cloning, pose risks ranging from compromised devices to system failure, impacting patient care and safety ~\cite{hamilton2018counterfeit, guin2014counterfeit}. To mitigate these risks and safeguard patient well-being, there is a critical need to develop techniques for detecting counterfeit devices within complex interconnected systems.  Authentication and identity management play crucial roles in uniquely identifying devices, enabling the detection of counterfeit ones. However, identifying counterfeit devices in complex CPS networks, such as those in IoMT and Telehealth, presents formidable challenges. Traditional compliance verification methods like visual inspections and documentation checks are slow, costly, and impractical, particularly in networks with remote devices ~\cite{guin2014counterfeit}.
\par
Hence, an innovative device identification approach is crucial to effectively detecting alterations like counterfeiting or cloning. In this context, we propose a framework to secure the integrity of medical images and imaging devices during image acquisition. By integrating techniques for device identification with medical image data provenance, our framework addresses the pressing need to combat device counterfeiting and cloning, ensuring the reliability and trustworthiness of medical systems across diverse applications like MCPS, IoMT, eHealth, and telehealth/telemedicine.

\subsection{Medical Image Data Provenance through Device Fingerprinting and Watermarking}
To solve the medical image data provenance problem, this study proposes a framework for securing the integrity of medical images using techniques such as DFP and image watermarking during the image acquisition process on these devices. Image watermarking the DFP makes it invisible to common users using mHealth-based devices in community healthcare services. The DFP derived from the imaging devices' attributes is embedded into the captured image via the image watermarking technique, ensuring the integrity and authenticity of subsequent tasks such as retinal disease diagnosis, screening, and monitoring. 
\par  
The proposed method uses software techniques to extract the device's features, generating a DFP that acts as payload data during the embedding process via image watermarking. Implementing these software applications demands computational, storage, and communication resources. However, it encounters limitations due to the inadequate computational, storage, and networking capabilities of legacy devices, and the quick replacement of these devices is unfeasible. Hence, in this study, we utilize our previous research "\textbf{DevFing: Robust LCR Based Device Fingerprinting}" from \cite{devFing2021Kumar}. This study introduced a unique solution employing an additional board called \textbf{DevFing}, which utilizes a daughterboard to capture the intrinsic characteristics of an electrical device, thereby generating a unique signature, termed Device Fingerprint (DFP).  Leveraging the electrical characteristic values of the device's electronic board, the daughterboard produces a stable DFP without compromising the device's primary functionalities. 
These signatures enhance security by effectively identifying and intercepting counterfeit devices. Finally, the study evaluates the proposed method of DFP generation and the medical image data provenance method's effects on image quality, computational overhead, and robustness to different image tampering-based attacks. 
This technology has the potential to revolutionize healthcare delivery, particularly in low-income areas with limited access to healthcare resources. Integrating these technologies into healthcare can improve the quality of care, increase efficiency, and reduce healthcare costs. Therefore, it is ideal for low-income users with limited healthcare facilities.
\par 

\subsection{Contributions}\label{subsec:contributation-medical image data provenance}
Key contributions are summarised as follows:

\begin{itemize}
    \item We propose a robust framework for tracking and verifying the authenticity of medical data, particularly fundus imaging datasets. Our framework enhances medical data management and ensures its reliable and secure use.

    \item To identify the source of medical images, this work uses DFP-based methods, which ensure the performance of the data for diagnosis. Furthermore, it used the unique features of the image to identify and investigate image tampering.

    \item This study utilizes the DWT-based image watermarking method to embed the source device's and image's unique information for their identification while implementing the proposed framework.


    \item This study analyses the effect of the proposed method of DFP-based watermarking in terms of various technical overloads, including computational time, image qualities, security, and trustworthiness.

     \item Finally, this study introduces a \textbf{DevFing}-based additional daughterboard for DFP generation for legacy devices without impacting its functionalities for developing and underdeveloped nations to implement the proposed medical image provenance framework in current fast-growing medical technologies and services.
    
\end{itemize}

\par
The remainder of the paper is organized as follows: Section ~\ref{sesc: related works}  highlights related studies on medical data provenience and source identification methods. Section ~\ref{sec:device-identification and Source} describes the proposed DFP and watermarking-based approach for medical image provenance framework in detail. Section~\ref{sec:Implementation and Results} provides results of various pipeline stages of the proposed framework. Finally, Section ~\ref{sec:conclusion} concludes the papers and discusses how the work can be extended in the future.

\begin{table}[t!]
\centering
\caption{Comparison of Data Provenance Techniques in Medical Imaging}
\fontsize{7.5}{8}\selectfont
\begin{tabular}{|p{0.09\linewidth}|p{0.22\linewidth}|p{0.22\linewidth}|p{0.28\linewidth}|p{0.06\linewidth}|}
\hline
Technique & Description & Advantages & Limitations &  Ref.\\
\hline


Metadata Annotations & Add detailed annotations to image metadata & -Widely Accepted, Basic Information, Rich Information, Customizability & Manual Effort, Annotations can be lost, May not track processing history  & \cite{ahmed2023data,gibaud2008dicom} \\ \hline

Water-marking & embedding an invisible digital watermark into the image & Resistant to common image manipulation techniques & Degrade the quality of the medical image data. May not track processing history & ~\cite{ahmed2023data,pan2009medical,nabi2022comprehensive} \\ \hline


Digital Signatures & A cryptographic technique apply digital signatures for authenticity &  Data Integrity,  Tampering Detection & Key Management Required, Doesn't track processing history & \cite{ahmed2023data,pan2009medical,nabi2022comprehensive} \\ \hline

Time-stamping & Record time-stamped logs of image-related events &  Chronological History, Real-Time Tracking & Storage Demands, May not capture fine-grained steps & \cite{johns2023data,pan2023data,ahmed2023data} \\ \hline

Secure Hashing & Calculate cryptographic hashes for integrity &  Efficient Integrity Checks, Applicability & Doesn't capture history,  Key Management Needed & \cite{johns2023data,ahmed2023data}\\ \hline



Blockchain & Use blockchain for immutable ledger &  Data Security,  Tamper Resistance &  Scalability Challenges,  Requires Blockchain Infrastructure & \cite{mcbee2020blockchain,margheri2020decentralised,sekhon2023extensive,d2022data} \\ \hline

Machine learning & ML techniques to analyse patterns in the data & Detect and prevent data tampering & Require a significant amount of training data, May not be effective against sophisticated attacks & \cite{pan2023data,ahmed2023data}\\ \hline  

\end{tabular}
\label{table:data-provenance}
\end{table}

\section{Related Works}\label{sesc: related works}
With the development of smartphones and Internet-enabled handheld devices, it will also be possible to acquire and share images and videos relating to a patient's health. This task is more challenging in developing nations such as India, where the patient-to-doctor ratio is low and healthcare availability is lacking. Many people need to be checked out in these situations, so social workers or health care agencies are used. However, healthcare and social workers committed fraud in this scenario by replacing original photographs with fabricated ones obtained through unauthorized software or hardware. A similar incident was reported recently in which a medical practitioner created a forged document related to COVID-19 treatment to make an insurance claim~\cite{fake2023MedicalReport}. Furthermore, in ~\cite{KumbhMe2021fakeCovid}, another similar incident was reported involving the generation of forged COVID-19 test reports for over 1 lakh people during the Kumbh Festival 2021 in Haridwar, India, by a private agency. Additionally, the consequences of fraud in healthcare applications are substantial, as they can impact people's lives.
Therefore, it is crucial to have a framework to find and classify fake and real images that a remote user shares with a centralized system or an expert ~\cite{ahmed2023data}.
\par

\subsection{Medical Image Data Provenance and Source Identification Frameworks}\label{subsec:device-identification and Source}
    The emergence of affordable and imaging-based portable devices has introduced new security vulnerabilities, especially in medical applications where data integrity directly impacts patient safety~\cite{johns2023data,bai2021security}. 
    Therefore, medical image data provenance is crucial in ensuring the authenticity and integrity of medical images. To address this issue, medical device identification information is typically incorporated into the metadata that accompanies medical photos. Metadata contains image-related information, such as the device used to capture the images, image resolution, pixel depth, time, location, and camera settings~\cite{johns2023data,pan2023data,ahmed2023data,gibaud2008dicom}. However, the metadata information of an image can be altered easily. Further, many techniques, such as digital signatures, watermarking, metadata, certificates, time-stamping, hashing, blockchain, and machine learning, have also been proposed for this purpose~\cite{johns2023data,pan2023data,ahmed2023data,margheri2020decentralised,bai2021security}. Table ~\ref{table:data-provenance} summarises and compares the various medical image data provenance approaches. In most of the data provenance approaches, the storage of security and data source information on device memory is a concern, as malicious users can easily access it in community setups. Moreover, blockchain offers a decentralized and tamper-resistant ledger but faces scalability challenges ~\cite{margheri2020decentralised}. ML can detect tampering patterns but requires substantial training data~\cite{pan2023data}. 
    Therefore, there is a need for a robust source identification method that does not rely on predefined or system-generated keys or identities stored in device memory. 

\par 
Therefore, to address this problem caused by local storage of device identification-related information, we proposed a new medical image data provenance framework that generates provenance information on the fly without storing it locally in the device directory. Additionally, this also ensures its integrity and suitability for low-resource setups.

\begin{figure}[t!]
    \centering
    \includegraphics[width=0.8\linewidth]{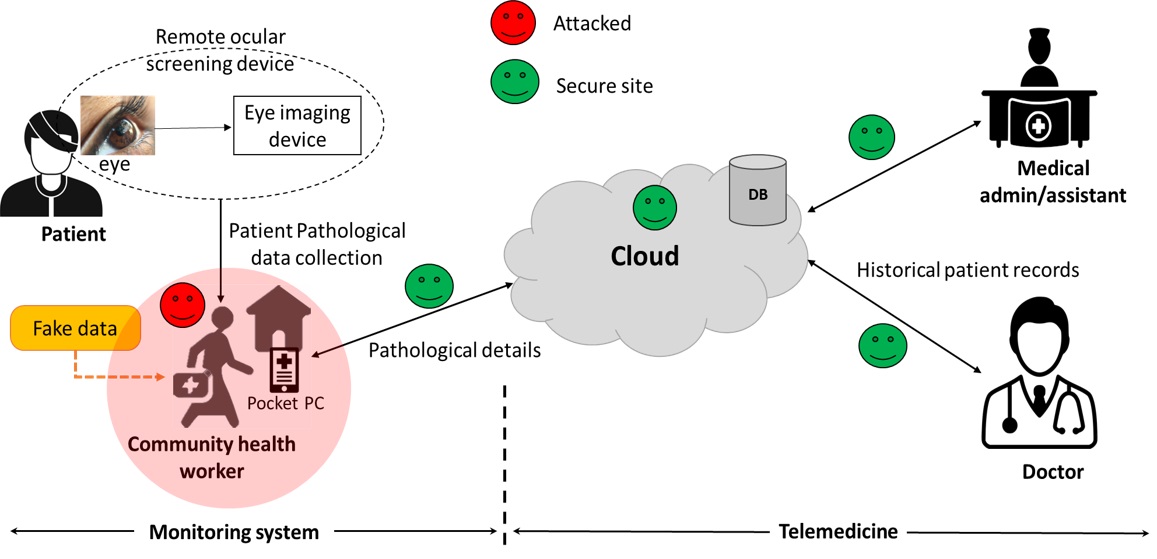}
    \caption{Attack model for fundus imaging-based affordable healthcare applications in IoT or mHealth. The red smiley symbol is used to show the potential attack site, and the green color smiley is used to depict the secure site in the proposed telecommunication setup.}
    \label{fig:mHealth application attack model}
\end{figure}

\section{Methodology: Medical Data Provenance Framework}~\label{sec:device-identification and Source}
This section introduces a framework for device identification and medical image data provenance in affordable healthcare systems such as telemedicine, eHealth and IoMT. It also explores relevant attack models, aiming to enhance the security and trustworthiness of imaging datasets.

\subsection{Threat Models}\label{subsubsec:attack model}\label{subsec: device provenance threat models}
\par 

This study is divided into two stages. As a result, we explore potential threats in each scenario separately.

\par 
\subsubsection{Compromised Operator Scenarios in Remote Healthcare System}
In the first stage, we assume that the imaging device (or healthcare device) operators have been compromised and have performed data tampering, manipulation, and fake image data. This assumption is particularly relevant in healthcare systems situated in remote locations with limited device security and surveillance. Under these circumstances, alternative methods for verifying the integrity and provenance of image data acquired and shared by operators in case of compromise are lacking. Consequently, within this system, we consider the following attack models (shown in Figure ~\ref{fig:mHealth application attack model}):
\begin{enumerate}
    \item We assume that all communication channels are secure. These channels are between users' local pocket personal computer (or other handheld devices) and the cloud, as well as between the cloud and the users (i.e., doctors, medical administration, medical assistants, and patients).
     
    \item All information (e.g., images, device characteristics) obtained by mobile healthcare devices or their applications is legitimate, and the acquisition process is entirely secure. 
    
    \item In the proposed attack model, we have considered the community health worker node is compromised, so it performs a fake injection or replaces medical image data related to the patient.  
    
    \item We assume that the system (mobile and handheld device) had sufficient security measures in place to detect malicious files or data contents, except for altered or fake images.

    \item In the proposed healthcare system, all nodes are secure except for the community health worker.
    
    \item Cloud-based operations such as data storage and processing, user authentication and validation, and information sharing, are secure and encrypted. 
\end{enumerate}

\begin{figure}[t!]
	\centering
	\includegraphics[width=0.6\linewidth]{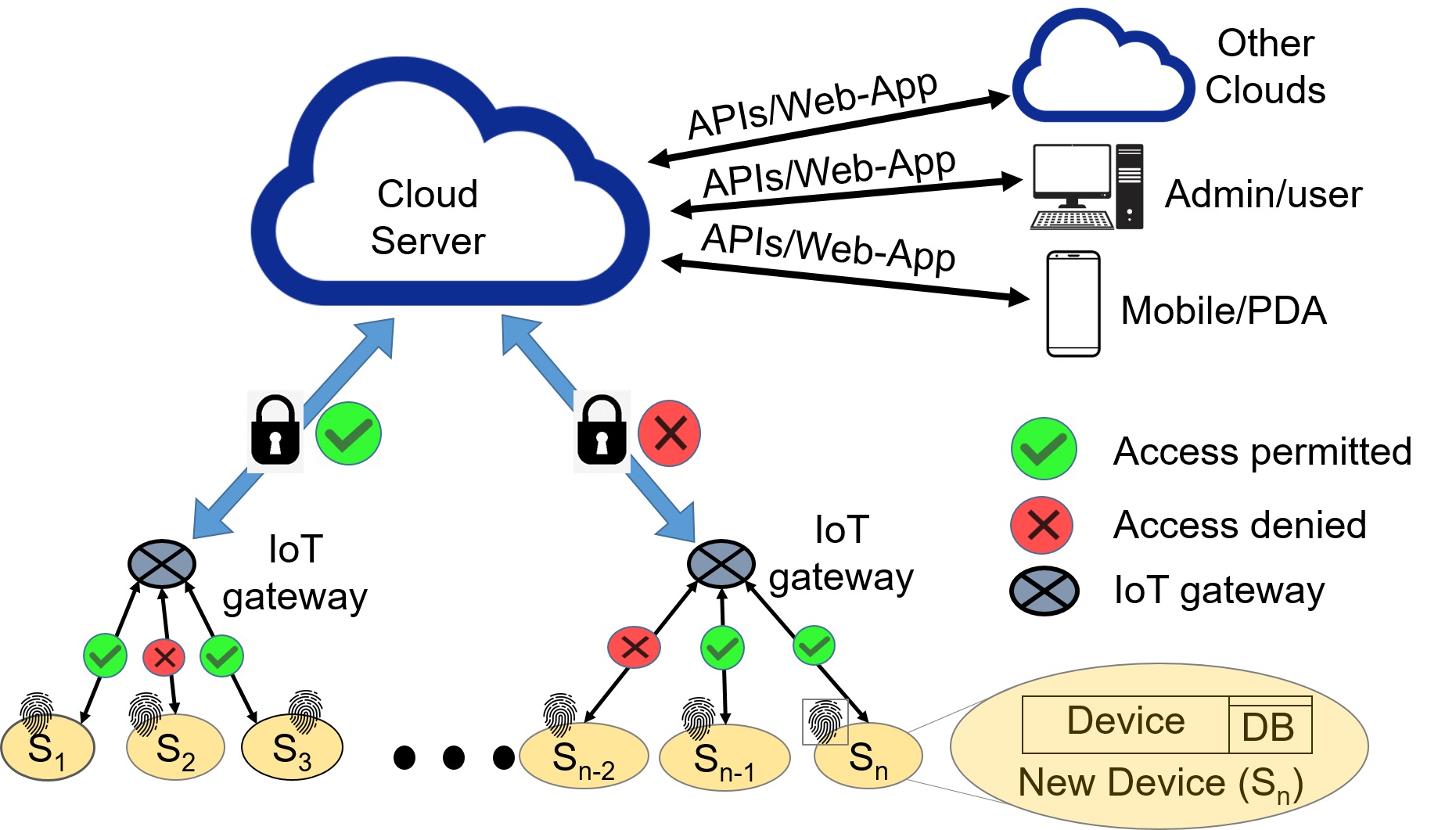}
	\caption{The device fingerprint application in IoT, IoMT, mHealth or telehealthcare system.}
	\label{fig:iot-application of device fingerprint}
\end{figure}

\subsubsection{Security Considerations for Legacy Devices in Remote Healthcare Settings}
In the second stage, we considered the case of legacy devices and the possibility of a healthcare device being compromised due to physical tampering or counterfeiting, especially in remote healthcare settings with limited physical security. This extension allows us to focus on safeguarding the integrity and authenticity of patient data while addressing the security implications of such compromised devices.  In this model, we assume the remaining system components are legitimate and secure while addressing the security implications of compromised devices. Therefore, we have also assumed (shown in Figure ~\ref{fig:iot-application of device fingerprint}) as part of the attack model:
\begin{enumerate}
    \item That our target system or device is secured against distributed denial of service (DDOS) or man-in-the-middle (or networking attacks) during any communication with the other device or server. All communications between the daughter board and test board are secure and encrypted to achieve security.
    
    \item The daughterboard is immune to side-channel attacks from an adversary for the device features collection, processing, and authentication using the board's power, timing, electromagnetic interference (EMI), and noise analysis.
    
    \item During authentication, we have considered only one device sending the unique device signature to the authentication server for self-authentication (or identification).
\end{enumerate}

The healthcare system is designed to provide a trustworthy and safe platform for distant healthcare services (i.e., telemedicine, mHealth, eHealth or IoMT, etc.) in the above-proposed attack models. It does this by leveraging advanced security measures such as DFP and watermarking to ensure the integrity and authenticity of medical data and early detection of compromised devices to prevent potential threats to security and privacy.
\par

\subsection{Proposed Framework for Medical Image Data Provenance}\label{subsec:proposed framework and its architecture-data provenance}
\begin{figure}[t!]
    \centering
    \includegraphics[width=0.85\linewidth]{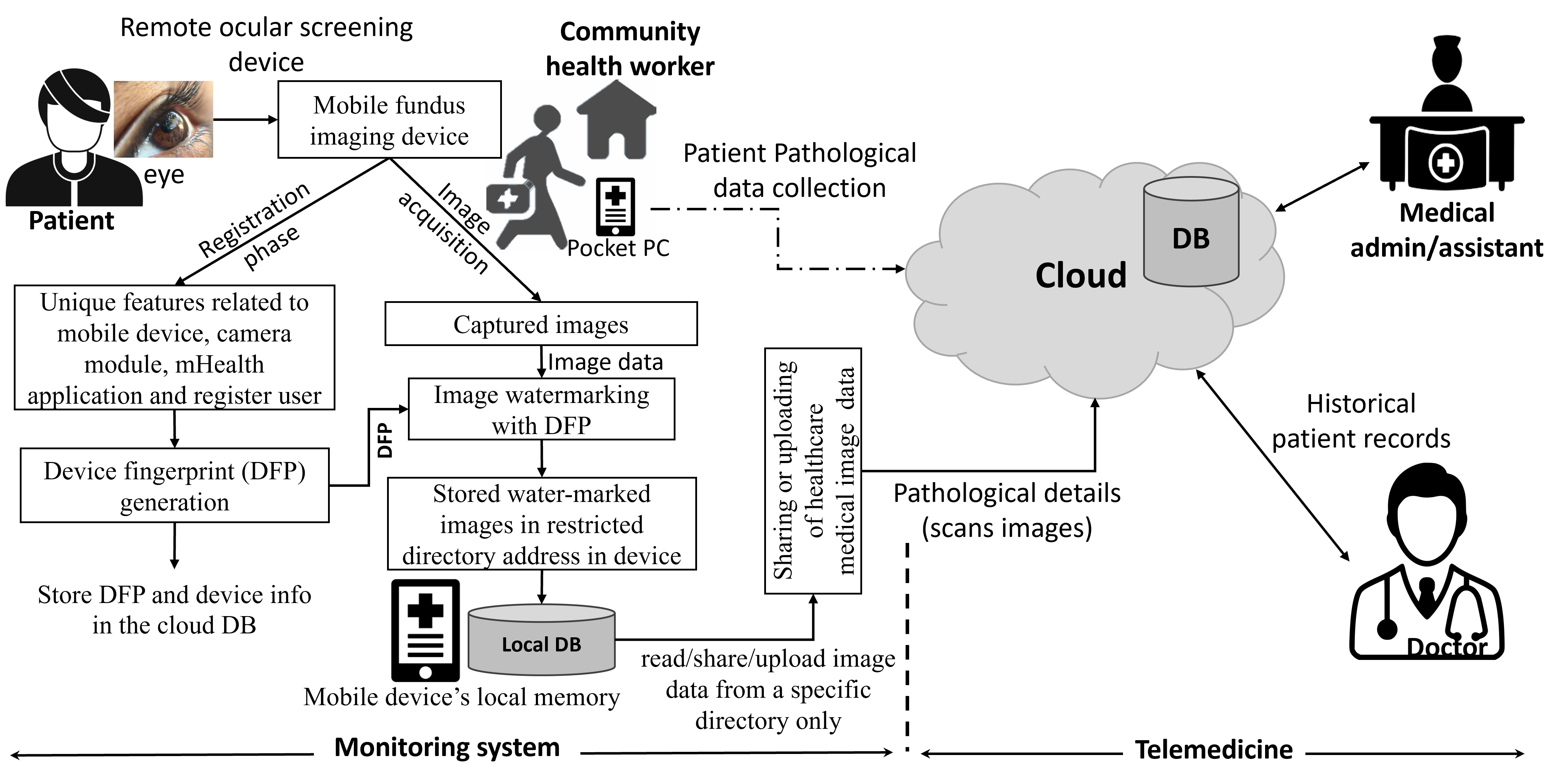}
    \caption{Architecture of proposed framework for image data provenance in mHealth or telemedicine applications. (The dotted line shows the old data flow diagram from Pocket PC to the cloud server directly without a provenance framework.)}
    \label{fig:mHealth security architecture}
\end{figure}

The detailed architecture of the proposed framework is shown in Figure ~\ref{fig:mHealth security architecture}. The framework has two working phases: (1) device registration and (2) image acquisition and sharing. In the registration phase, a new medical device with its unique identity is registered on the cloud-based medical or healthcare network. Here, we have used the concept of DFP-based device identification. DFP is a function of the device's hardware and software features~\cite{vijay-devicefingerprint-acm2023}. After successfully registering the device on cloud-based systems, a copy of the DFP (i.e., the device's unique signature) is stored for feature referencing and identification during the validation/authentication phase. 


\par 
The provenance framework consists of the following functional units: DFP generation unit, image acquisition and watermarking unit, image file handling unit, image source information retrieval unit, and authentication unit. In the subsequent sub-sections, we explore the functionalities of each unit individually.

\subsubsection{Working of Provenance Model}\label{subsubsec: working of provenance model}
\begin{figure}[t!]
    \centering
    \includegraphics[width=0.8\linewidth]{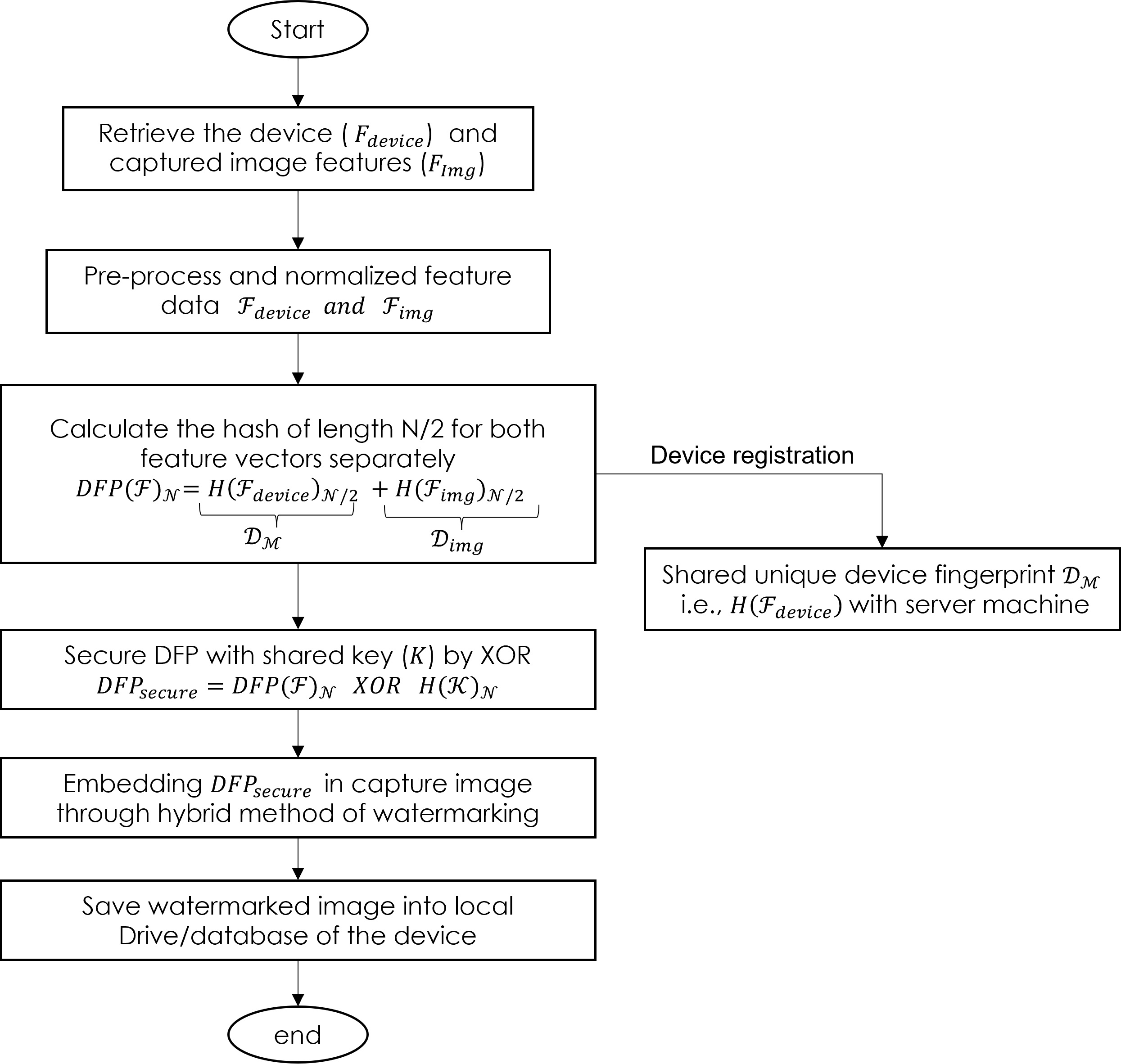}
    \caption{Flowchart illustrating the Generation of DFP for Device and Image, along with the device registration procedure}
    \label{fig:device registration and DFP genration}
\end{figure}

Figure ~\ref{fig:device registration and DFP genration} illustrates the functioning of the proposed framework for DFP generation and device registration in the proposed provenance model on a local device. In this, the framework captures images through an imaging unit and embeds source and image information using image watermarking before storing it in a local directory. The framework employs the device-unique DFP for source identification and utilizes the image's unique signature for image identification. Furthermore, the framework restricts the application's capability to read, share, and upload images solely from a specific directory on the mobile device's local memory (where watermarked images are stored during acquisition) to the cloud (refer to Figure ~\ref{fig:mHealth security architecture}). Finally, during data exchange, the cloud-based system evaluates image data quality and source to authenticate and validate the imaging device and received images. This approach is designed to prevent and identify instances of image data alteration and the utilization of compromised, unregistered devices.

\subsubsection{DFP Generation Unit}\label{subsec:Device fingerprint generation unit}
A DFP generation unit creates a unique identifier ($\mathcal{D}_{M}$ or signature) for device ($M$) identification. This unit consists of two sub-systems: devices' feature extraction and fingerprint generation. The feature extraction unit captures and extracts the device's distinctive features ($F$) to achieve a unique device identification. Subsequently, fingerprint generation units utilize these features to generate a unique signature for device identification. Two distinct systems are proposed for this purpose. Firstly, an application-based system is suggested, where the device's application software extracts information about its hardware, software, running applications, services, and user details as unique feature sets. 
\par
Furthermore, consider a scenario where the device itself is compromised. In such instances, modern devices equipped with advanced communication, data storage, and processing capabilities can be remotely monitored through various application software. This monitoring allows the acquisition of real-time hardware and software statuses and hardware configurations. However, replacing all outdated devices quickly is challenging in developing or underdeveloped countries. As a result, numerous medical devices in such conditions lack the necessary resources to support computation, storage, networking, and data-sharing capabilities. To solve this problem, we propose an extra daughterboard in the second system that can capture the device's electronic board's electrical characteristics without impacting its essential functions and generate a unique signature. Section ~\ref{subsec:device fingerprint genration} of this study provides a detailed illustration of both DFP generation methods.
\begin{figure}[b!]
    \centering
    \includegraphics[width=0.7\linewidth]{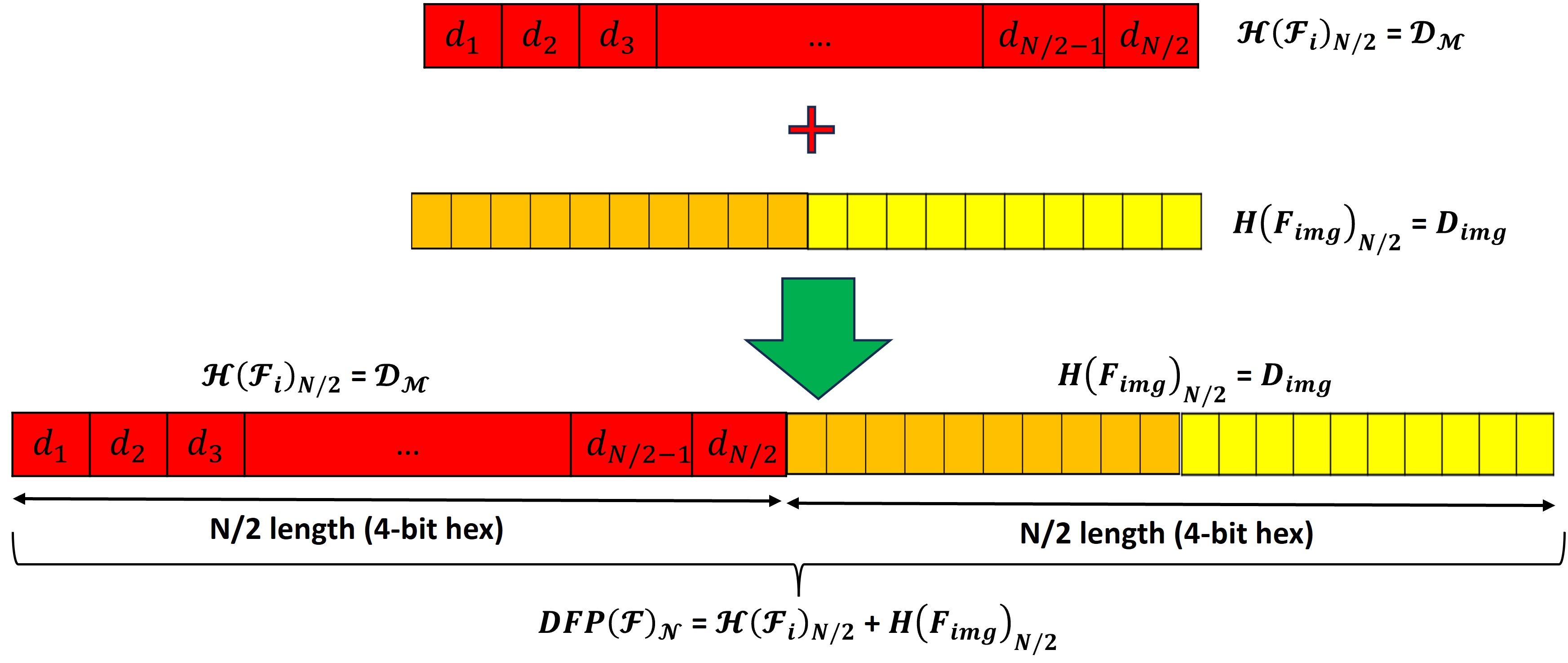}
    \caption{Generation of a unique device fingerprint using a mobile device and image information.}
    \label{fig:Generation of a unique Device Fingerprint}
\end{figure}


\subsubsection{Image Acquisition and Watermarking Unit}\label{subsubsec:image acqusition and watermarking unit}
This unit captures new images and extracts their unique signature ($\mathcal{D}_{img}$,) to represent the newly captured image. The signature generation unit utilizes spatial variation in an image to create a unique signature for image identification. $\mathcal{D}_{img}$ is a hexadecimal number string of length equal to $\frac{N}{2}$ used to uniquely represent the image.  
\par 
\par
\begin{figure}[t!]
    \centering
    \includegraphics[width=0.85\linewidth]{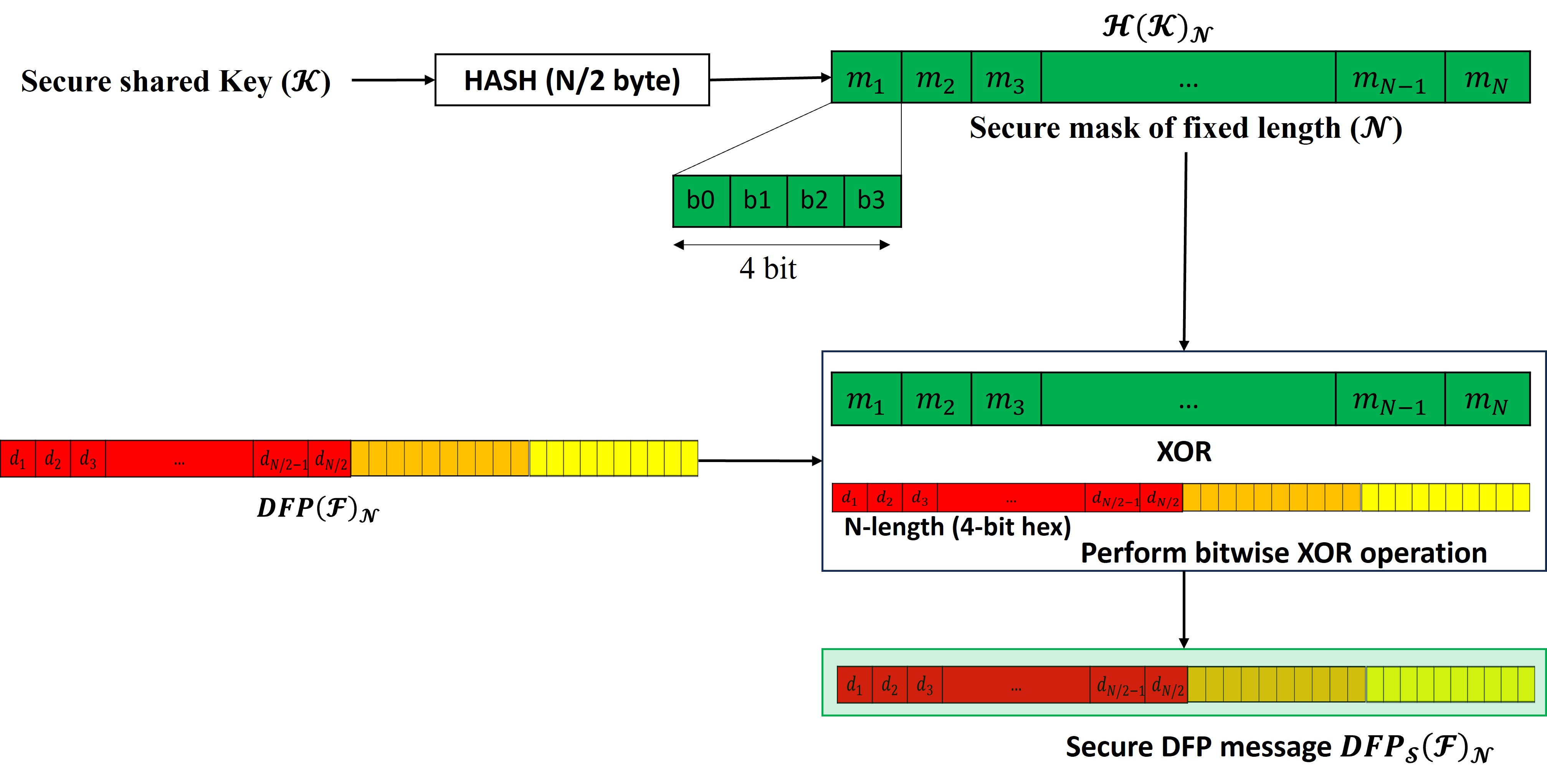}
    \caption{Secure DFP before embedding it in the image using pre-shared key information.}
    \label{fig: process to secure a DFP before embedding}
\end{figure}

\begin{figure}[t!]
    \begin{minipage}[c]{0.46\linewidth}
        \centering
        \includegraphics[width=01\textwidth]{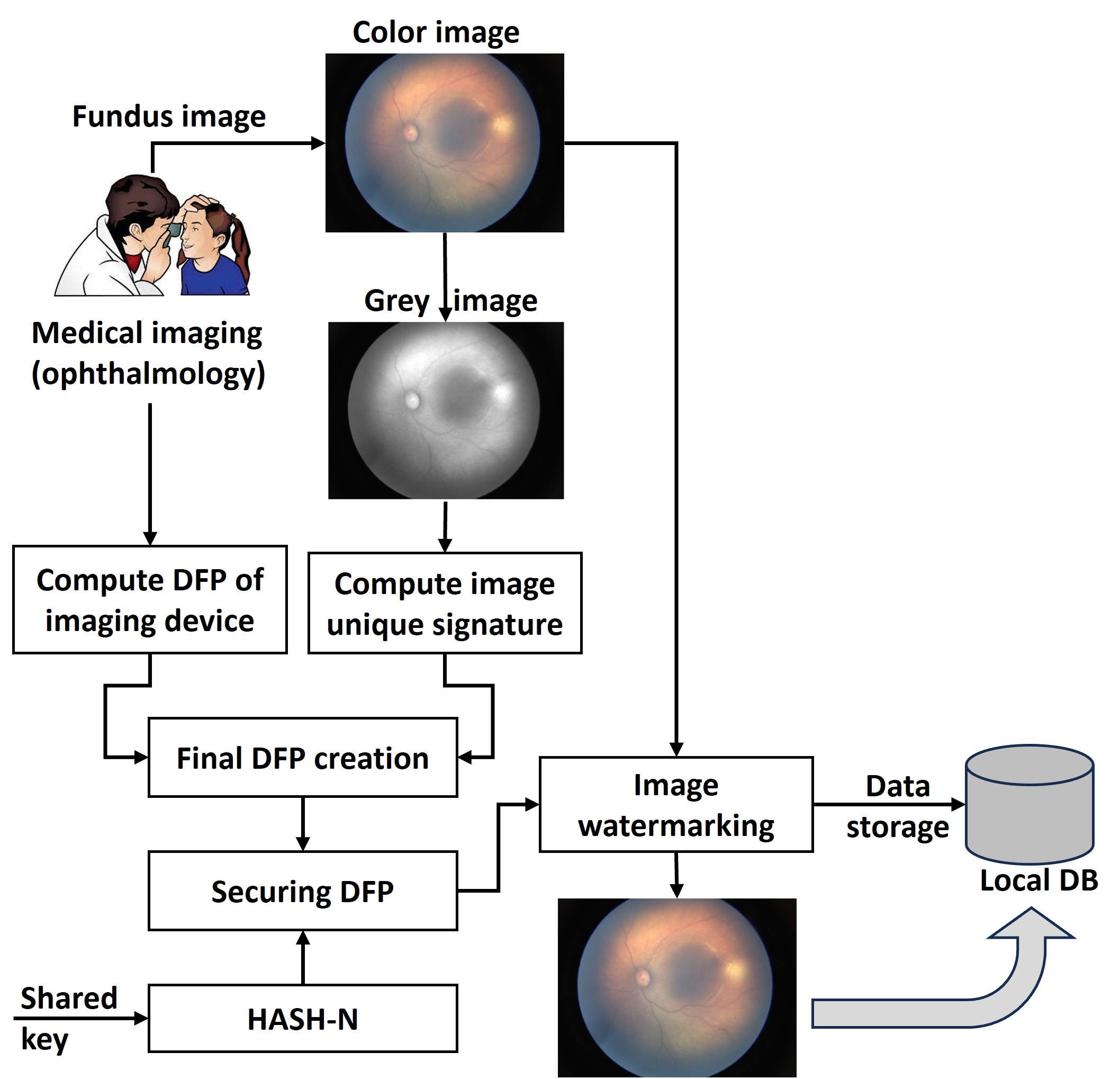}
        \caption{Secure image watermarking and storage in a local database.}
        \label{fig:secure image watermaking and image storage}
    \end{minipage}
    \hfill
    \begin{minipage}[c]{0.5\linewidth}
        \centering
        \includegraphics[width=01\textwidth]{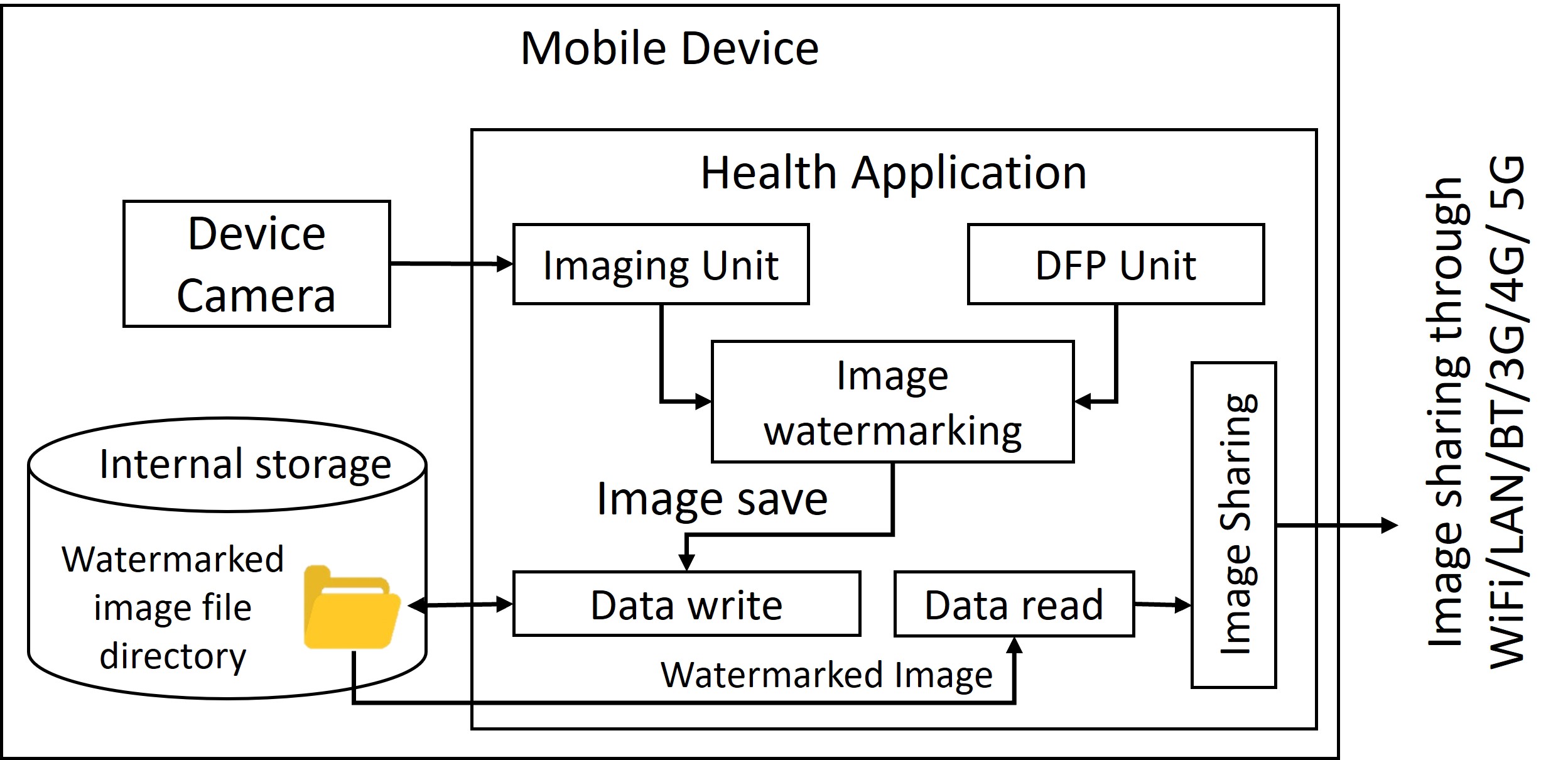}
        \caption{Working of image file handling unit.}
        \label{fig:mobile device file handling}
    \end{minipage}

\end{figure}

 
After generating the image signature, the device fingerprint embedding unit combines the image signature and device fingerprint, securing it with pre-shared key information. The details of this procedure are illustrated in Figure ~\ref{fig:Generation of a unique Device Fingerprint}. This process involves concatenating the image and the device's unique information side by side, forming a longer hexadecimal number stream ($DFP(\mathcal{F})_{\mathcal{N}}$) with a length equal to $\mathcal{N}$. Subsequently, it secures the newly generated hexadecimal stream with a secure mask with a bit-wise $\mathcal{XOR}$ operation. This mask is generated by employing fixed-length hashing of the pre-shared key ($\mathcal{K}$). The complete process of securing DFP information before embedding it into a captured image is shown in Figure ~\ref{fig: process to secure a DFP before embedding}. After computing a secure DFP representing the device and the image, this unit embeds this information into the captured image using an image watermarking algorithm.
The mathematical expression for embedding the DFP into the captured image is given by $ \mathcal{I_{D}} = \mathcal{I} + \alpha * \mathcal{D}_{M} $, where $\mathcal{I_{D}}$ is the image with the embedded device fingerprint, $\alpha$ is the scaling factor determining watermark strength, and $+$ is various watermarking techniques that can be applied. These techniques include Least Significant Bit (LSB), Intensity Shift Bit (ISB), patchwork, Discrete Cosine Transform (DCT), Discrete Fourier Transform (DFT), Discrete Wavelet Transform (DWT), Singular Value Decomposition (SVD), DWT-SVD, etc.~\cite{gull2023advances}. Several recent approaches have explored the utilization of ML/DL-based techniques for image watermarking. However, it is noteworthy that these methods require considerable resources for the development of DFP embedding and extraction models~\cite{wan2022comprehensive}. Following this, the watermarked images are stored within the local directory of the device, reserved for subsequent applications. Figure ~\ref{fig:secure image watermaking and image storage} provides a visual depiction of the comprehensive process, covering the acquisition of medical images, the incorporation of secure information through watermarking, and the subsequent storage of these processed images in a predefined directory on the local device for future applications.

\subsubsection{Image File Handling Unit}\label{subsec:Watermarked image uploading and sharing unit}
Figure~\ref{fig:mobile device file handling}, illustrates the image file handling system architecture and working. 
This unit handles file management operations related to writing,  reading, sharing and saving operations.
In Figure ~\ref{fig:mobile device file handling}, the image file handling unit facilitates the storage of watermarked images ($\mathcal{I_{D}}$) in the local file system of the device. The data-sharing unit then accesses these watermarked images exclusively for subsequent sharing with remote databases, servers, clouds, or experts, enabling consultation, management, disease screening, diagnosis, and analysis. These units play a critical role in regulating the data flow by granting selective access to files and data (i.e., images) in a designated storage location within the device, where only uniquely watermarked images reside. In the proposed framework, both the storage location for watermarked images and the functioning of the data-sharing unit remain fixed and the same, ensuring secure and controlled data transfer.

\subsubsection{Image Source Information Retrieval and Authentication}\label{subsubsec:Cloud based image source extraction and validation unit}
After receiving medical image data from a remote mobile device, the cloud-based system initiates a verification process using the cloud application. This process involves extracting the source information (i.e., imaging device) and image details from the received data. Once this information is extracted, the system verifies the authenticity of the imaging device or source. This step ensures that only authenticated and registered devices can capture images. Therefore, the system automatically detects such incidents if a malicious user attempts to use a new device.
\par
\begin{figure}[t!]
    \centering
    \includegraphics[width=0.9\linewidth]{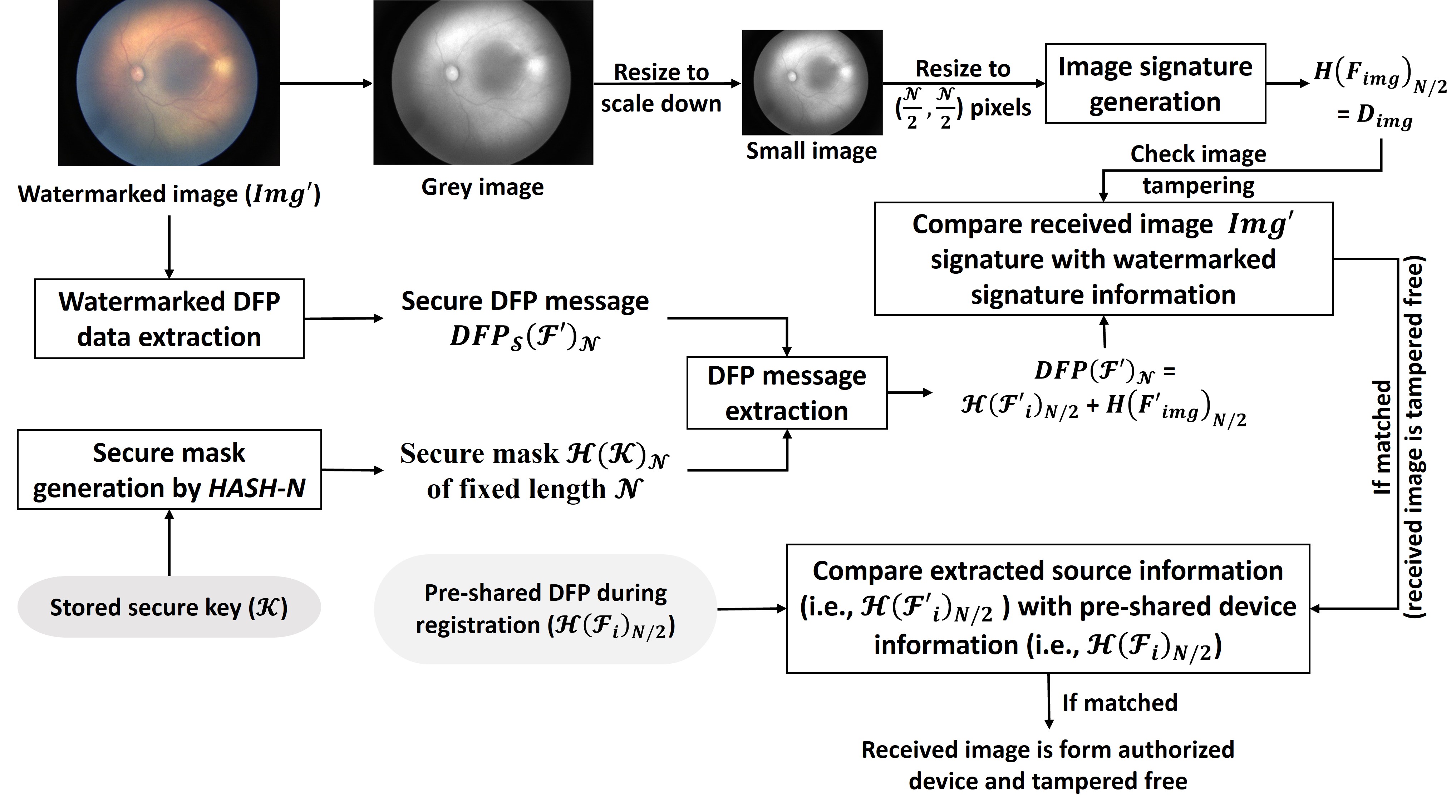}
    \caption{Extraction of device source information and unique image signature, followed by identification of image tampering and authentication of the imaging device's source}
    \label{fig: extraction of secure watermarked information and  authentication}
\end{figure}

 
Furthermore, the system utilizes the extracted image features to authenticate the received image. A detailed description of the stages, including source and image feature extraction and the authentication of the imaging device and image, is illustrated in Figure~\ref{fig: extraction of secure watermarked information and  authentication}.
\par 
The extraction unit is crucial in telemedicine setups, providing source information authentication for received watermarked images. This involves extracting the unique device identification information ($\mathcal{D}'_{M}$) from the watermarked image ($\mathcal{I_{D}}$), which may be affected by various artefacts, noises, and attacks. The extracted information ($\mathcal{D}'_{M}$) is then compared with pre-stored source information ($\mathcal{D}_{M}$) for image authentication. Remote applications can then utilize authenticated images for medical diagnosis and consultation.

\begin{figure}[t!]
    \centering
    \includegraphics[width=0.9\linewidth]{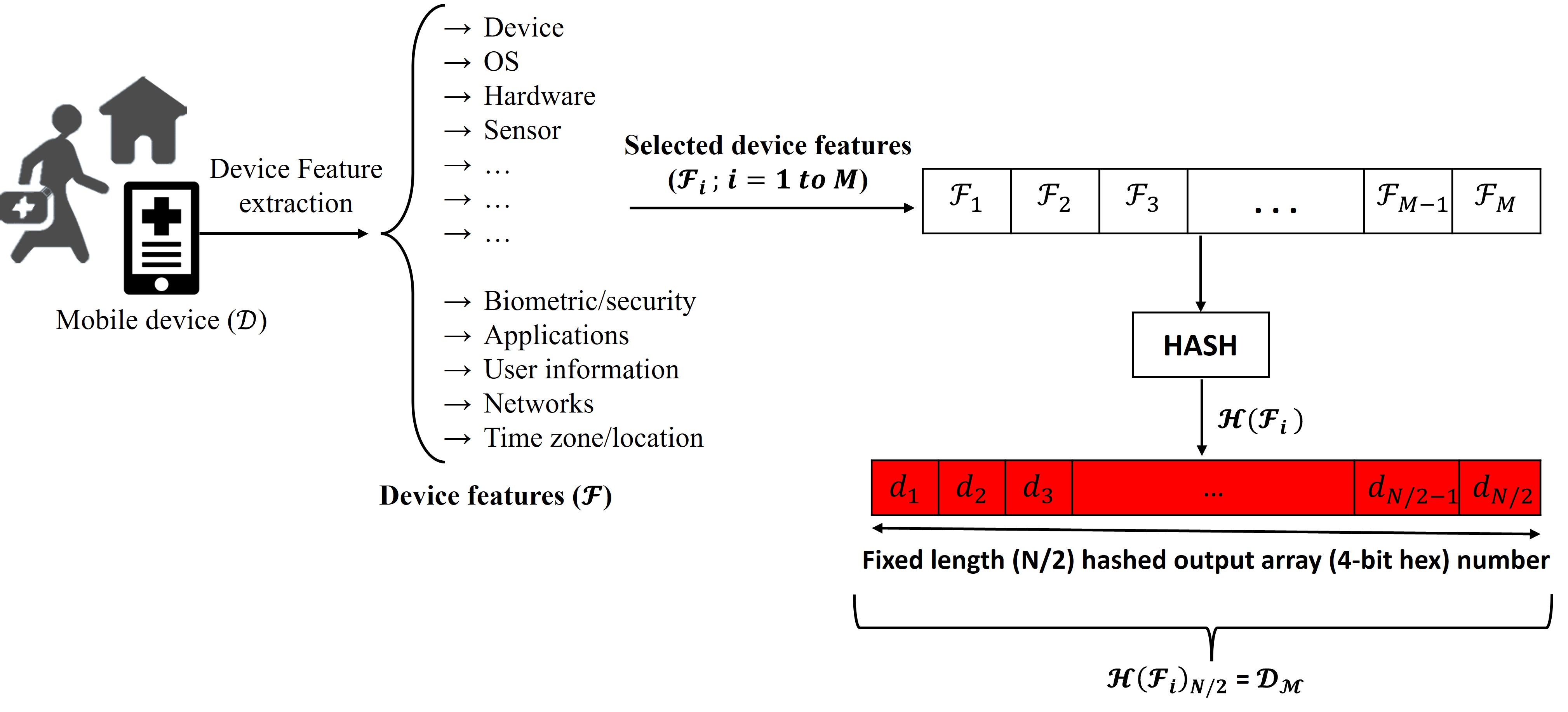}
    \caption{Mobile device feature extraction and fixed length device hash creation.}
    \label{fig:mobile device feature extraction and hashing}
\end{figure}

\begin{table}[b!]
    \centering
    \fontsize{8}{7}\selectfont
    \caption{Various mobile device features ($\mathcal{F}$) used in device fingerprint generation.}
    \begin{tabular}{|p{0.25\linewidth}|p{0.7\linewidth}|} \hline
    Feature type & Device feature \\ \hline

    Basic device information & Device manufacturer, Device model, Operating system version, Kernel version, Build number, Language, Time Zone, etc. \\ \hline
       
    Hardware information & CPU architecture, Number of CPU cores, Amount of RAM and ROM, Internal storage capacity, Battery percentage, Boot time   \\ \hline
    
    Sensor information & Type and number of sensors available, Sensor resolution and accuracy\\ \hline
       
    Camera information & Camera resolution, Number of cameras, Flash type  \\ \hline
       
    Screen information & Screen resolution, Screen density, Screen size  \\ \hline
       
    Touchpad information & Type of touchpad, Number of touchpads, Resolution  \\ \hline
       
    Biometric information & Fingerprint sensor availability, Iris sensor availability, Face detection  \\ \hline

    Application information & List of installed applications \\ \hline
    
    User information & User account name, UID    \\ \hline
    
    Network information & Type of network module (WiFi/BT/4G/5G), MAC IDs, Mobile Number, IMEI, IP address, Gateway   \\ \hline

\end{tabular}
\label{tab:list of mHealth device features}
\caption*{UID: Unique Identifier, IMEI: International Mobile Equipment Identity, IP: Internet Protocol, BT: Bluetooth}
\end{table}

\subsection{Application-based DFP Generation}\label{subsec:device fingerprint genration}
In an application-based DFP generation mode, the DFP generation unit uses software algorithms that uniquely identify and acquire the device's characteristics to generate a unique signature to identify it.

\subsubsection{Basic Concept and Background}
Over time, DFP techniques have evolved into an essential method for device identification and detecting malicious activities~\cite{ibrahim-2019-formalization}. DFPs find widespread application in ICT and Cyber-Physical Systems (CPS), addressing various aspects such as security, hardware and software fault detection, user behaviour analysis, identity verification, and more~\cite{vijay-devicefingerprint-acm2023}. Specifically, when considering the hardware attributes of handheld medical imaging devices, elements like the central processing unit (CPU), random access memory (RAM), read-only memory (ROM), sensors, camera type, image resolution, communication modules, etc., show stability over time~\cite{vijay-devicefingerprint-acm2023}. Furthermore, details about the device's software, containing the operating system (OS), software versions, installed applications, and system services, play a key role in generating a unique DFP. Additionally, user-specific characteristics, such as preferences, interaction patterns, and application usage history, further contribute to the distinctiveness of the device signature. Collectively, these features form the intrinsic identity of the device, remaining relatively consistent over time. Consequently, in an application-based system, the device application software extracts information related to the device's hardware, software, running applications, services, and user details as unique feature sets, facilitating the generation of a DFP for device identification and management.

\subsubsection{Architecture and Working of DFP Generation}
This section discusses the application-based DFP generation module. It has two sub-modules: \textbf{device feature extraction} and the\textbf{ DFP generation unit}, as depicted in Figure ~\ref{fig:mobile device feature extraction and hashing}. 
\paragraph*{\textbf{Device Feature Extraction Unit: }}
Particularly for handheld imaging-based medical devices ($\mathcal{M}$), the feature extraction unit extracts the various characteristics ($\mathcal{F}$) of a device, such as its hardware and software configurations, and combines them into a unique identifier. In detail, this unit retrieves all relevant device characteristics, such as its make and model, operating system version, processor type, and other hardware and software configurations, which are listed in Table~\ref{tab:list of mHealth device features}.

\par
For device feature extraction, an API can be employed to obtain real-time data from the device. Specifically, in mobile phones operating on either Android or iOS-based systems, numerous built-in APIs are accessible for application development. These APIs serve the purpose of collecting and acquiring real-time data associated with the device's usage, facilitating further processing and device monitoring. In the context of this DFP generation model, software applications collect this information and generate a unique signature for device identification. Details of the API packages used in Android applications are described in \href{https://developer.android.com/reference/android/hardware/fingerprint/package-summary}{android/hardware/fingerprint/package-summary}. Further, in \href{https://github.com/fingerprintjs/fingerprintjs-android}{fingerprintjs-android}, developers employ these packages to create a lightweight library for device identification and fingerprinting. Similarly, the library is also available for browser fingerprinting as well as iOS systems at \href{https://github.com/fingerprintjs/fingerprintjs-ios}{fingerprintjs-ios} and \href{https://github.com/fingerprintjs/fingerprintjs}{fingerprintjs} respectively.

\paragraph*{\textbf{DFP Generation Unit: }}
Further, the second unit processes the retrieved device features to generate a stable DFP. Once the device information ($\mathcal{F}$) has been retrieved, it needs to be normalized to ensure that the fingerprint remains consistent across different devices. This involves removing unnecessary information, such as spaces or special characters, and converting all data to a common format.  Further, it arranges all extracted features into a standard, predefined sequence. Further, it employs $SHA256$ hashing algorithms to digest this normalized processed feature sequence to generate a fixed-length unique sequence.

\begin{figure}[b!]
    \centering
    \includegraphics[width=.8\linewidth]{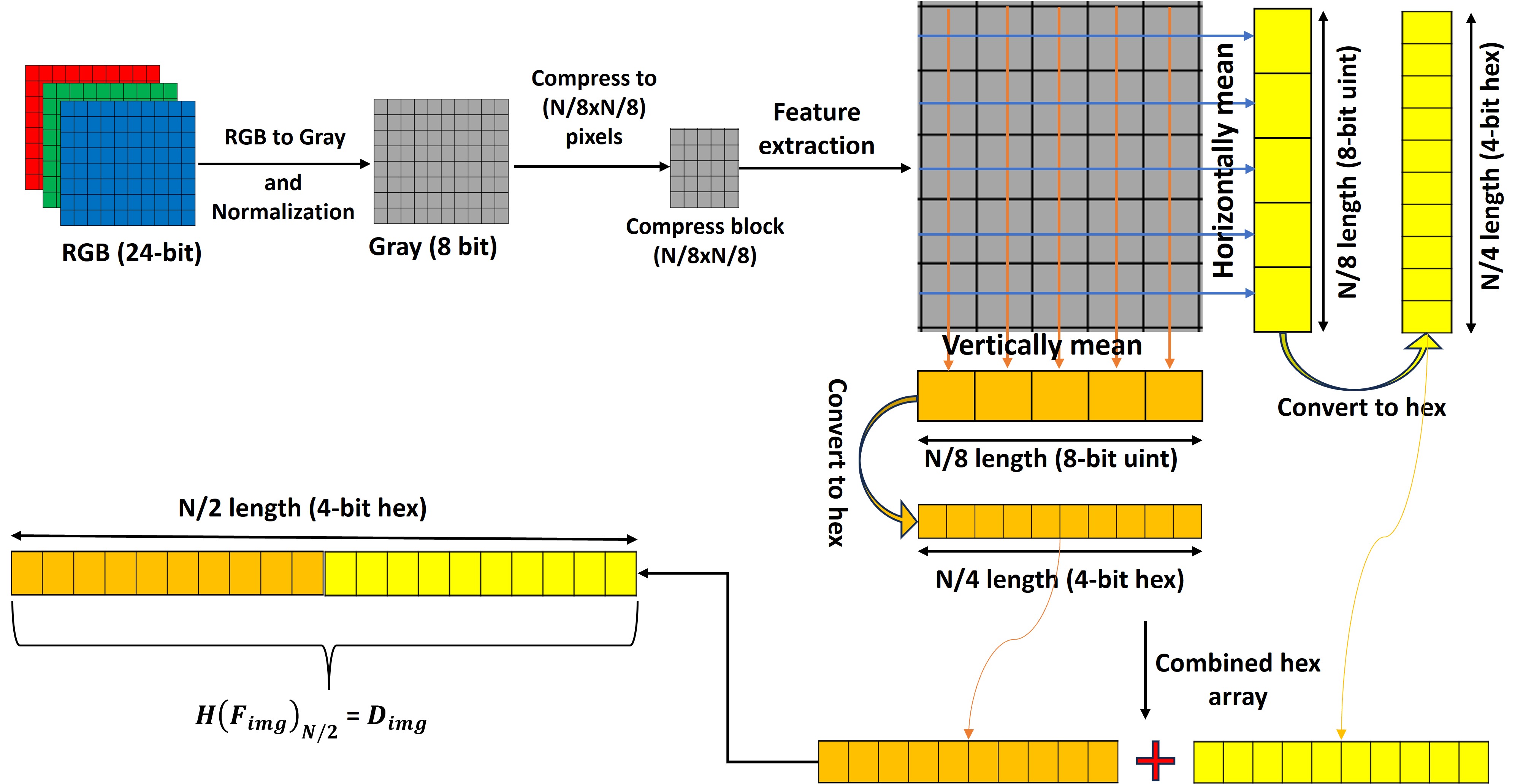}
    \caption{Image feature extraction for unique image identification: illustrating the generation of a unique hexadecimal array sequence through spatial averaging of grey intensity in compressed images, both horizontally and vertically.}
    \label{fig: Image feature extraction process}
\end{figure}
\subsubsection{Working of Image Signature Generation}
The image signature encapsulates the spatial irregularities within an image. This computation involves the calculation of spatial component averages in two directions: horizontally and vertically. The resulting data from both the horizontal and vertical directions is then merged into a single sequence, effectively doubling its length. Before the amalgamation with the device signature ($\mathcal{D_{M}}$), the proposed approach needs the conversion of this Unit8 array into a hexadecimal array, also with double the original length, to facilitate the fusion with the device fingerprint array information. 
\par 
The workings of the image signature generation method are illustrated in Figure ~\ref{fig: Image feature extraction process}. 
In this process,  it converts a color image with a 24-bit depth to an 8-bit grayscale image, normalizing its spatial intensity values between 0 and 255. Subsequently, it reduces the grayscale image to a size equal to $\frac{N}{8} \times \frac{N}{8}$ (where N = 256) and computes the spatial average in both horizontal and vertical directions. Afterwards, it converts both Uint8 sequences into a hexadecimal number sequence of length $\frac{N}{4}$. The following step combines both hexadecimal sequences to generate a hexadecimal number sequence of length $\frac{N}{2}$. This $\frac{N}{2}$ length hexadecimal number sequence, represented by $\mathcal{D}_{img}$, is then used to represent the image uniquely.

\par 

Figure ~\ref{fig: Image feature extraction process},  provides a visual explanation of the complex process of extracting unique features from an image to enable its distinctive identification. The central concept revolves around the spatial averaging of grey intensity within a compressed image, conducted separately in horizontal and vertical directions. The resulting values are meticulously compiled into a unique hexadecimal array sequence, forming the basis for the image's identification.
\par
We next discuss \textbf{DevFing}, a mechanism that can create a DFP for older devices with limited computing, storage, and communication capabilities. 

\subsection{Hardware-based DFP Generation}\label{subsec:DevFing architecture}
 In this section, we discuss the architecture of our proposed daughterboard for reliable fingerprint extraction.


\subsubsection{Background and Device Electrical Intrinsic Characteristics of a Device} \label{subsubsec:background and electrical features for DFP genration}
Particularly in the era of IoT, system technologies are used to advance very quickly. Therefore, the system architecture is also evolving at the same pace, which requires frequent software updates and upgrades. Due to this, the device parameters (such as clock skew, delay, power consumption, etc.) also change quickly.
Over the past few years, many methods and frameworks for tracing and identifying the source of medical data have been made. To overcome the device identification, in this direction, Van Der Leest et al. have proposed a Hardware Intrinsic Security (HIS) mechanism that can provide security based on intrinsic properties of the device~\cite{van2013anti}. The HIS-based techniques are suitable for device identity and authentication because they are unclonable, unpredictable and easy to fabricate ~\cite{herder2014physical,Chatterjee2017PUF}.

\par
During the manufacturing process of the printed circuit board (PCB), the copper traces, epoxy layer, soldering of components, chipsets, etc., have some structural or compositional variation of materials. These variations are reflected in their electrical or electronic properties. Moreover, in the multi-layer PCB, these variations are increased due to the heterogeneous and complex structure. In every electronic device, the PCB (Printed Circuit Board) possesses unique electrical characteristics, including inductance (L), capacitance (C), resistance (R), and impedance (Z)~\cite{hamilton2018counterfeit}. Therefore, in \cite{wei2015boardpuf}, authors used unique intrinsic properties such as capacitance and inductance of the board to identify a PCB. In ~\cite{zhang2015robust}, authors exploited trace impedance variations of the board for PCB identification. However, these PCB-based techniques are prone to operating conditions such as temperature, humidity, electromagnetic noises, etc.; ageing and device tampering may cause the bit flip error~\cite{wei2015boardpuf}. In ~\cite{paley2016active}, authors proposed a new technique that provides active protection against PCB physical tampering by exploiting the board's resistor and capacitance. The effect of board/circuit ageing cannot be ignored, as it affects all components of a circuit board, including PCB traces, mounted components, and chipsets, causing the board's performance to deteriorate over time. Recently, in ~\cite{quadir2018low} and ~\cite{martin2020notchpuf}, authors exploited the PCB electrical characteristics in the identification of electrical filters such as low-pass and Notch filters. However, in this, the authors don't acknowledge the effect of the ageing of the board on the performance of a DFP.
\par
To mitigate the above problem, we proposed a new system that reduces the effect of ageing on device fingerprinting. It leverages the electronics board's L, C, R, and Z values to generate a reliable DFP. The proposed system uses an additional board (a daughterboard) with the mainboard. The daughterboard will perform all functional tasks needed for signature generation and authentication.

\subsubsection{Detailed Design of the Device Signature Generator}\label{sec:System architecture}
\begin{figure}[!t]
\centerline{\includegraphics[width=01\linewidth]{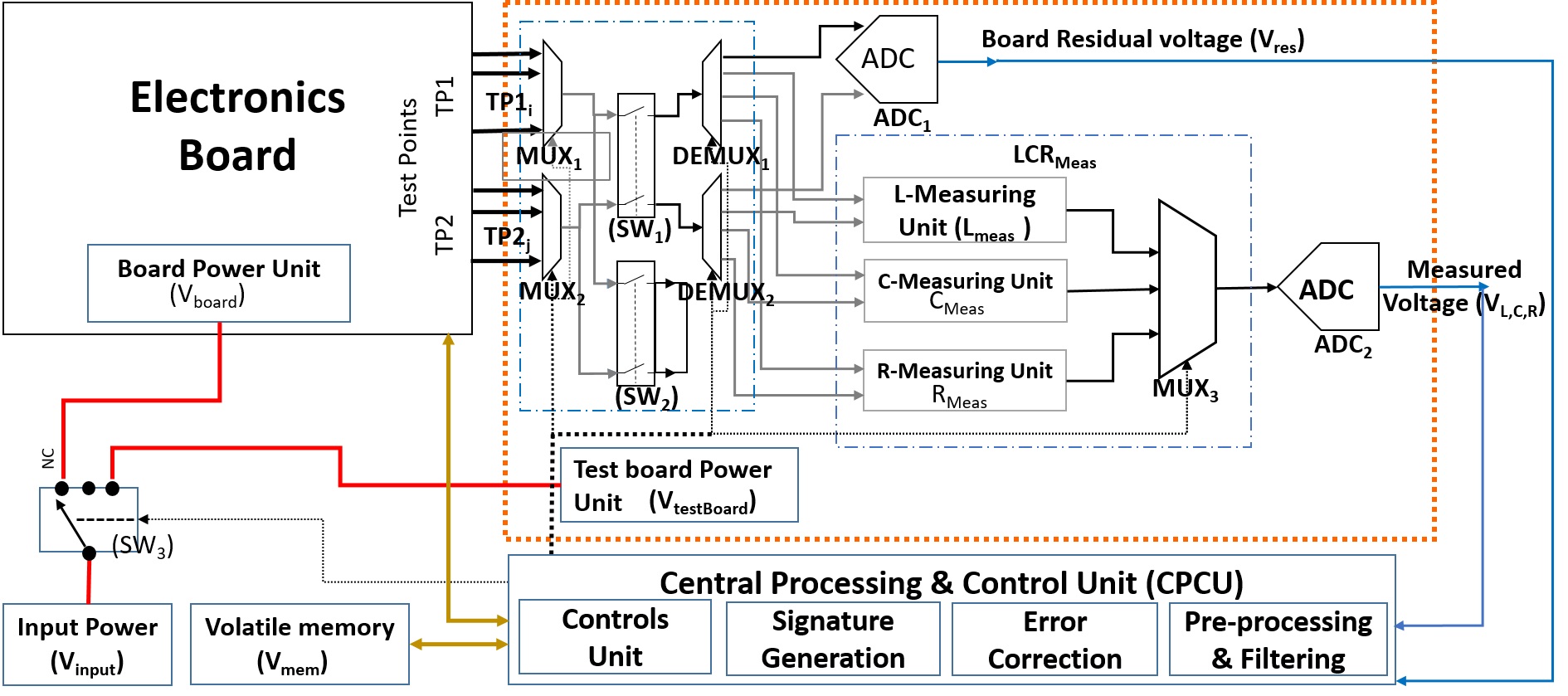}}
\caption{Anatomy of the device signature generator.  It uses the board's intrinsic characteristics, such as inductance, capacitance, and resistance, to generate the device signature.}
\label{fig:System architecture}
\end{figure}

In this section, we discuss the architecture of our proposed daughterboard for reliable fingerprint extraction. Before discussing the functional architecture and workings of the system, we consider the following:
\begin{itemize}
	\item  The electronics board has sufficient test points in the two sets (TP1 and TP2) to measure the LCR value, as shown in Figure ~\ref{fig:System architecture}.
	\item  Test points in each set are denoted by $TP1_{i}$ and $TP2_{j}$, respectively, and $i$ and $j$ range from $0$ to $N-1$, where $N$ is the total number of test points. 
\end{itemize}{}

The functional architectural description of the system is shown in Figure ~\ref{fig:System architecture}, which comprises the following functional entities: 
\begin{itemize}
\item \textbf{Electronic Board and Test Points: }  
The electronic board is the test device whose electronic signature needs to be calculated. The multiple test points are being selected on the PCB to measure the board's LCR values. These test points are chosen so that they are homogeneously spread across the complete board.  To get the representative test points, we have used a simple uniform random test point selection method~\cite{Leskovec2006sampling}.

\item \textbf{Test Point Selection Unit: }
The test-point selection block (inside the dotted box) is a compact network of the multiplexer (MUX), demultiplexer (DMUX), and electronic switches, as depicted in Figure ~\ref{fig:System architecture}. $MUX_{1}$ and $MUX_{2}$ are connected to the two sets of test points  $TP1$ and $TP2$ respectively. These $MUX$s serve the purpose of selecting the pair of input test points each from both test point sets $TP1$ and $TP2$. $MUX$ operation is controlled with the help of the $MUX$ control signal, which is connected to the processing and control unit. $MUX_{1}$ and $MUX_{2}$ outputs are being given to both electronic switches $SW_{1}$ and $SW_{2}$, where $SW_{1}$ provides a soft, controlled bridge between the measurement units and a selected pair of signals and $SW_{2}$ output connects with two DEMUX blocks (viz., $DMUX_{1}$ and $DMUX_{2}$). DEMUX blocks have four outputs connected to the ADC block $ADC_{1}$ and LCR measurement unit $LCR_{meas}$.

\item \textbf{LCR Measurement Unit: } 
LCR measurement unit ($LCR_{meas}$) comprises three electrical parameter measurement units, namely, the inductance ($L_{meas}$), capacitance ($C_{meas}$) and resistance ($R_{meas}$) measurement units. Each measurement circuit is used successively to generate highly precise output voltages $V_{L}$, $V_{C}$, and $V_{R}$  for L, C, and R, respectively, across the test pair. However, these measurement circuits use existing standard L, C, and R measurement techniques with predefined operating frequencies ($f$ in Hz). The measured output voltage is routed through the $MUX_{3}$ to provide a stable voltage output for the digitization in the ADC($ADC_{2}$) block at a time.

\item \textbf{Power Management Unit: } Power Management Unit ($P_{mgmt}$) consists of three separate power units to control the proposed electronic system and the board power. These include the Board Power Unit ($P_{board}$), the Test Board Power Unit ($P_{testBoard}$), and the System Power Input Unit ($P_{input}$).

\item \textbf{Central Processing and  Control Unit (CPCU): } The CPCU is an external microprocessor or microcontroller unit responsible for controlling and generating board signatures. $CPCU$ provides a centralized platform to monitor and synchronize the board's acquisition of electrical parameters and unique signature generation.

\item \textbf{Volatile Memory: } The Volatile Memory ($VMem_{1}$) unit stores the unique device signature at run-time. Thus, the volatility may protect sensitive information (i.e., device signature), as it becomes unavailable once power is down. Volatile memory should be writable only from the daughterboard, and it must also be tamper-proof-free and immune to any fault.
\end{itemize}

\begin{figure}[!t]
	\centering
	\includegraphics[width=01\linewidth]{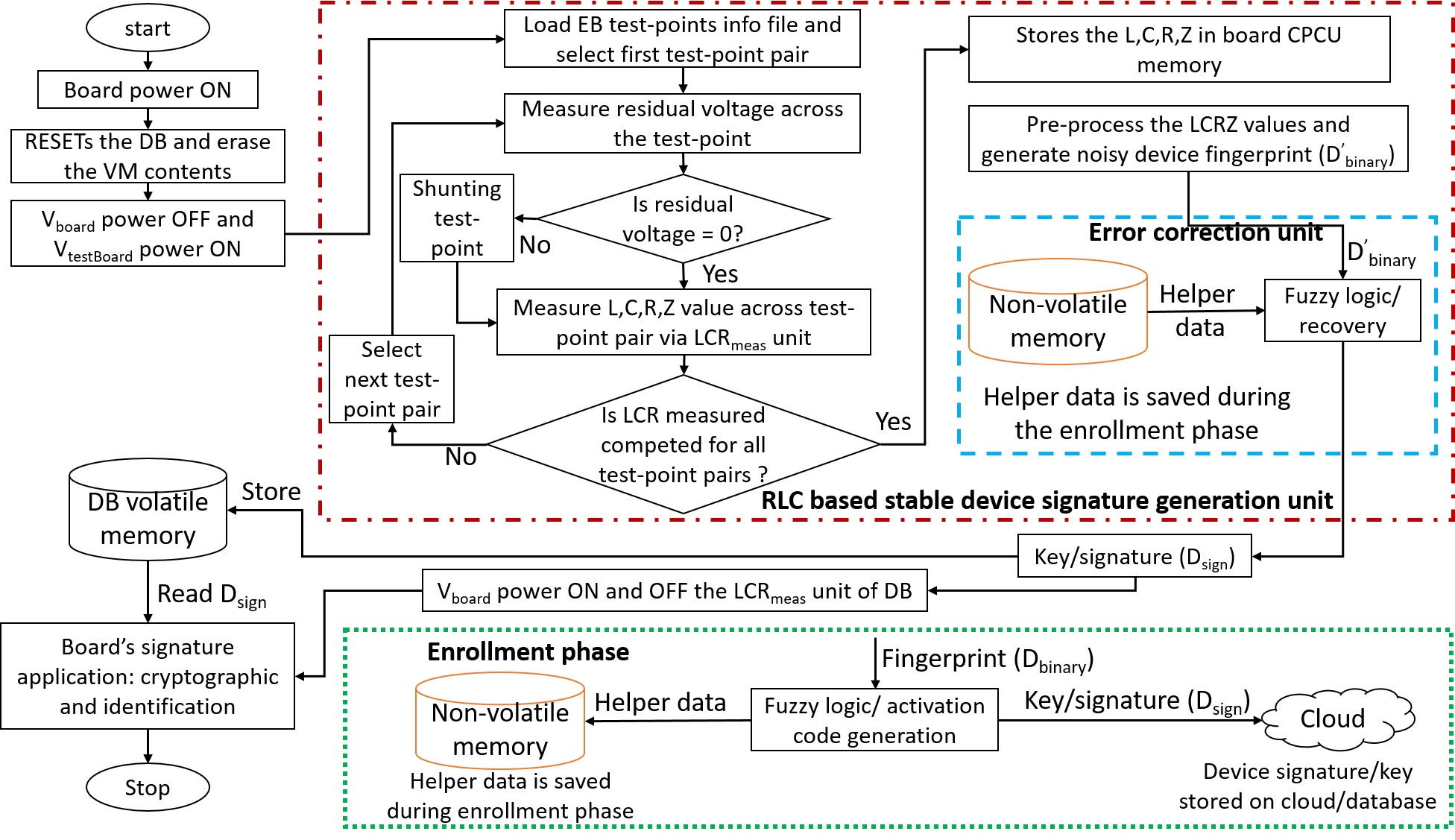}
	\caption{Flow chart: LCR-based stable DFP generation. (EB: electronic board, DB: daughter board and VM: volatile memory)}
	\label{fig:fingerpring system working flow chart}
\end{figure}

\subsubsection{Working of Device Signature Generation}
Figure ~\ref{fig:fingerpring system working flow chart} shows the workflow for \textit{DevFing} unique signature generator to generate a stable device signature. It employing the concept of PUF to generate a unique device identification based on electrical characteristics. It has three components:
\begin{itemize}
	\item[i.] \textbf{Board Power Management:} It regulates the power supply to various units on the board, namely, the Electronic Board Unit, LCR Measurement Unit, CPCU, and Volatile Memory Unit.
	\item[ii.] \textbf{Intrinsic Characteristic Based Signature Generator:} It measures the LCRZ values of the device and processes them to generate a stable device signature. 
	\item[iii.] \textbf{Signature Storage:} This unit stores the device signature for future applications.
\end{itemize}
 

\subsubsection{Device Feature Collection}
The device feature collection module measures the internal feature of the device and generates fingerprints from it. In this process, the \textit{DevFing} reads the L, C, R and Z values and stores it in the volatile memory. For all test and simulation purposes, we take board readings, and their corresponding L, C, R and Z values. The characteristics matrix ($ D_ {LCR}  = [L, C, R, Z]'$) for $N$ test point is:
\begin{equation}
D_{LCR} =\left[
\begin{matrix}
L_{0}&L_{1}& ... &L_{N-2}&L_{N-1}\\
C_{0}&C_{1}& ... &C_{N-2}&C_{N-1}\\
R_{0}&R_{1}& ... &R_{N-2}&R_{N-1}\\
Z_{0}&Z_{1}& ... &Z_{N-2}&Z_{N-1}
\end{matrix}
\right] 
\end{equation}
where, $L_{0}, ..., L_{N-1}$; $C{0}, ..., C_{N-1}$; $R_{0}, ..., R_{N-1}$ and $Z_{0}, ..., Z_{N-1}$ are the LCRZ value across $N$ test points.

\subsubsection{Fingerprint Generation}
The measured device device's electrical characteristics are still in raw format. Hence, data have various challenges, such as high dynamic range, stability, and measurement errors. The following steps to process the data to generate a fingerprint:

\begin{itemize}
	\item{\textbf{Step I}: } Takes the $log$ of all feature vectors to decrease their dynamic ranges. Originally, the feature sets, viz. L, C, R and Z have large values of their respective dynamic ranges. For example, the C values lie between $\mu F$ or $(10^{-6}F)$ to $pF$ or $ (10^{-12}F)$. Therefore, C has a high dynamic range, which is more than $10^{6}$. Similarly, the L, R and Z values of the board have also a high dynamic range.  The new value of L, C, R, and Z after $log$ operation can be expressed as $D_{log}$. 
\begin{equation}
    D_{log} = log(D_{LCR}) = [log(L), log(C), log(R), log(Z)]^{'}
\end{equation}\label{equ:D_log}

\item{\textbf{Step II}: } Take the mean (or average) of the logarithmic of the  R, L, C and Z values. Therefore, 
\begin{equation}
        mean(D_{log}) = [mean(log(L)), mean(log(C), mean(log(R), mean(log(Z)]^{'}
\end{equation}

\item{\textbf{Step III}: } 
Binarization of a device characteristic parameters involves converting them into a unique binary sequence. This process utilizes an algorithm that compares the calculated logarithmic values with the average logarithmic value for the corresponding characteristic values. Thereafter, we get four separate streams of binary data, i.e. $L_{binary}$, $C_{binary}$, $R_{binary}$ and $Z_{binary}$, respectively for $L$, $C$, $R$ and $Z$ sequences. Finally, these streams of data are considered as the device's signature or fingerprint.  For high accuracy, stability and uniqueness, we have taken the tuple of a binary sequence as a unique $DFP$. This is in the form of a tuple of the binary stream of {$L_{binary}$, $C_{binary}$, $R_{binary}$} and $Z_{binary}$ and called the device fingerprint $DFP$ (or $D_{binary}$). This is successfully presented in Algorithm ~\ref{algo:Intel Galelio Board Fingerprint Generation using DFP}.  
\end{itemize}

\begin{algorithm}
    \caption{DFP generation. The feature vector size: $m$ is the number of feature points, and $n$ is the number of test points.}
    \label{algo:Intel Galelio Board Fingerprint Generation using DFP}
    \textbf{Input:} $(D_{LCR})_{m \times n}$: Board's feature vector of size the $m \times n$. Where $m$ is the number of features, and $n$ is the number of test points. \\
    \textbf{Output:} $(D_{sign})_{n \times m}$: Binary stream of device signature with size of $n \times m$.    
    \begin{algorithmic}[1]
        \Procedure{DeviceSignGen}{$D_{LCR}$}
            \State $D_{log} \gets \log(D_{LCR})$
            \State $D_{binary} \gets (D_{log}^{T} \geq \text{mean}(D_{log}^{T}))$
            \State \textbf{return} $D_{binary}$
        \EndProcedure
    \end{algorithmic}
\end{algorithm}

\begin{figure}[b!]
	\centering
	\includegraphics[width=0.6\linewidth]{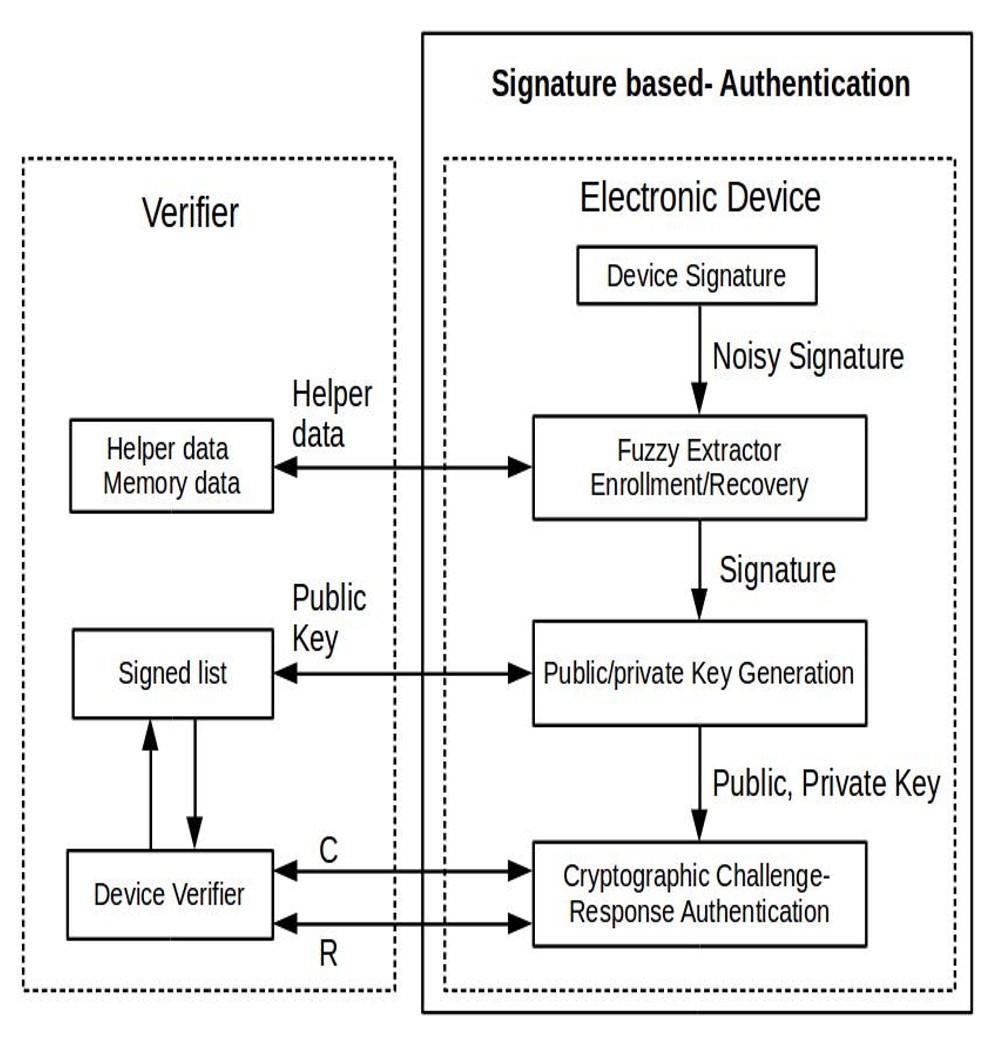}
	\caption{Block diagram of a device authentication method using the device signature. In this enrollment phase recovering the signature/key from the noisy signature.}
	\label{fig: Bolock diagram of a Device authentication method using signature}
\end{figure}

\subsubsection{Fingerprint Matching and Error Correction}\label{subsec:error correction}
In the proposed system, the \textit{DevFing} first generates the main board's signature which is based on their intrinsic characteristic parameters of a new device. Whenever a new device tries to connect with a network, it goes through the proper authentication process to identify genuine devices. In the proposed system, the pre-shared key (PSK) cryptography-based authentication process has been used, where the authentication block uses the device signature to generate the stable public private key for its authentication and validation. 
\par
The common error which appears in device fingerprinting due to ageing and tampering is a bit flip ~\cite{herder2014physical,wei2015boardpuf}. In IoT (or IoMT) devices, it arises due to harsh operating conditions, aging and device tampering. To overcome and mitigate this error, the proposed solution uses a fuzzy extractor based technique for recovering the signature during the enrollment or authentication process. The fuzzy extractor outputs are the key and helper data ($H$) string, which are used for error-correction. In the reproduction procedure, the input will be correctly constructed with small error (i.e, bit errors) only if the hamming distance between original and noisy information is less than the length of the hamming distance ($ham\_err$)~\cite{dodis2004fuzzy}.
\par

In our proposed system, the correction processes use the device features $D_{binary}$  and pre-stored helper bits ($H$), as shown in Figure ~\ref{fig:fingerpring system working flow chart}. The generation of a helper bit from the device signature is known as the enrollment phase. In the enrollment phase, a copy of helper bits is stored in a new device and a copy of the generated unique key ($D_{sign}$) is stored on the cloud-based database for further device authentication. The enrollment phase comes just after the manufacturing of a device but before releasing it. After release, the device is authenticated with the pre-stored cloud-based key. For authentication, it uses new DF ($D^{'}_{binary}$) (which is potentially corrupted because of the temporal, spatial and ageing) and are being compared with the golden fingerprint of the same device in the cloud-based DFP database (shown in Figure ~\ref{fig: Bolock diagram of a Device authentication method using signature}).

\section{Implementation and Results}\label{sec:Implementation and Results}
This section presents the implementation of the proposed framework and assesses its overhead costs, including computational complexity, image quality, and imperceptibility, specifically for imaging-based disease screening and monitoring. The framework's evaluation was conducted on a Dell Latitude 3400 laptop equipped with an Intel (R) Core(TM) i7-8565U CPU@1.80GHz and 16GB of RAM. We investigated the framework's impact on image quality using various datasets, listed in Table~\ref{tab: datasets used for medical image data provenance}, 
considering its potential application in imaging-based telemedicine setups. Further, for the evaluation of the hardware-based device fingerprint generation system DevFing, we have used two IoT-standard hardware: the Intel Galileo and the Raspberry Pi-3B.
\begin{table}[b!]
    \centering
    \caption{Dataset used for testing and validation of medical data provenance framework. }
    \resizebox{1\linewidth}{!}{%
    \begin{tabular}{|c|c|c|c|c|} \hline
        Dataset & Image Resolution & Image type & Image Depth & Disease/classes  \\ \hline
        Fundus image (IRIR dataset) &  1600x1200 & PNG & RGB & ROP, ROP-Plus, Healthy \\ \hline
        
        \href{https://github.com/Shenggan/BCCD_Dataset}{Blood cell}  & 640x480 & JPEG & RGB & Red Blood Cell, White blood cell, Platelets \\ \hline
        
        \href{https://www.kaggle.com/datasets/shubhamgoel27/dermnet}{Dermne Acne} & 720x480 & JPEG & RGB & Acne and rosacea ~ \\ \hline
        
        \href{https://www.kaggle.com/datasets/shubhamgoel27/dermnet}{Dermne hair loss} & 720x472 & JPEG &  RGB & Hair loss \\ \hline
        
        \href{https://www.kaggle.com/datasets/shubhamgoel27/dermnet}{Dermne-nail-fungus} & 467x720 & JPEG & RGB & nail fungus\\ \hline
        
        \href{https://www.kaggle.com/datasets/shubhamgoel27/dermnet}{Dermne-Actinic} & 720x474 & JPEG & RGB & Actinic \\ \hline
        
        Wound-foot-ulcer~\cite{wound-2021-dataset} & 512x512 & JPEG & JPEG & foot ulcer \\ \hline
        
        Skin-lesion~\cite{tschandl2018data} & 1022x767 & JPEG & RGB & Common pigmented skin lesions \\ \hline

        Monkeypox Skin Lesion & 224x224 & JPEG & RGB & Chickenpox, Measles \\ \hline
    \end{tabular}
    }
    \label{tab: datasets used for medical image data provenance}
\end{table}

\subsection{Application-based System for Device Fingerprinting}\label{subsubsec: results device fingerprint generation}
This section presents the analysis results of the proposed provenance framework, derived from various stages within it. This analysis assumes that medical imaging devices can perform application-based feature extraction and DFP generation.
\par
Therefore, we chose a smartphone as our example medical imaging device to validate the proposed framework. We utilized the smartphone's hardware and software characteristics as device features to generate a unique device fingerprint. Additionally, we considered multiple images from different datasets to analyze the effect on the quality of received images, overheads, and tampering detection during DFP embedding and extraction through image watermarking.

\subsubsection{Evaluation Metrics and Datasets}\label{subsubsec:evalution metrics-dfp based watermarking}
To evaluate the impact of DFP on the quality and accuracy of medical images after embedding, we used a set of eleven metrics, including mean squared error (MSE), root mean squared error (RMSE), peak signal-to-noise ratio (PSNR), RMSE of sliding windows (RMSE\_SW), universal quality index (UQI), structural similarity index (SSIM), error relative global adherence (ERGAS), spectral correlation coefficient (SCC), relative average spectral error (RASE), spectral angle mapper (SAM), multi-scale structural similarity index (MSSSIM), visual information fidelity (VIFP), and PSNR with blocking effect (PSNRB)~\cite{Wang2004QualityAssessment}. These quality assessment metrics allowed us to evaluate various aspects of our proposed approach, including computational time, image qualities, security, and trustworthiness. For the implementation of these image quality metrics, we utilized the Python package \href{https://github.com/andrewekhalel/sewar}{\textbf{Sewar}}.  

%
For testing and validating our proposed framework, we have carefully chosen multiple medical image datasets listed in Table~\ref{tab: datasets used for medical image data provenance}. Our selection criteria primarily focused on the potential future applications of mobile health devices in the screening, monitoring, and diagnosing of selected diseases. Consequently, in this study, we did not include images obtained from X-rays, CT-scans, magnetic resonance images (MRIs), ultrasounds, and other modalities that involve devices for a comprehensive evaluation of disease conditions. These modalities typically involve expensive, bulky, and complex equipment and the need for trained operators to perform the examinations.


   

   
   
   
   





\subsubsection{Fingerprinting of Mobile Healthcare Devices}
Smartphones and tablets have gained popularity in healthcare applications due to their efficient capabilities for acquiring, processing and sharing medical images and sensor data. To identify a mHealth device, we utilise a combination of device features, containing hardware and software attributes along with user details, to craft a unique signature known as DFP. These attributes were chosen from the list provided in Table~\ref{tab:list of mHealth device features}. We utilised the \textbf{hashlib} cryptographic module in Python to generate a device signature, implementing the SHA-256 algorithm. SHA is a cryptographic hashing function that generates a fixed-length output (HASH) based on input data, providing a unique digital signature for data integrity and authenticity. The resulting hash value serves as a device fingerprint, enabling accurate and efficient identification of the source device. Additionally, a fixed-length mask is generated after hashing the pre-shared secure key ($\mathcal{K}$), which is used to secure the watermarked message in the image.

\begin{figure}[b!]
	\centering
	\begin{subfigure}{0.3\linewidth}
		\centering
        \includegraphics[width=1\linewidth, height=4cm]{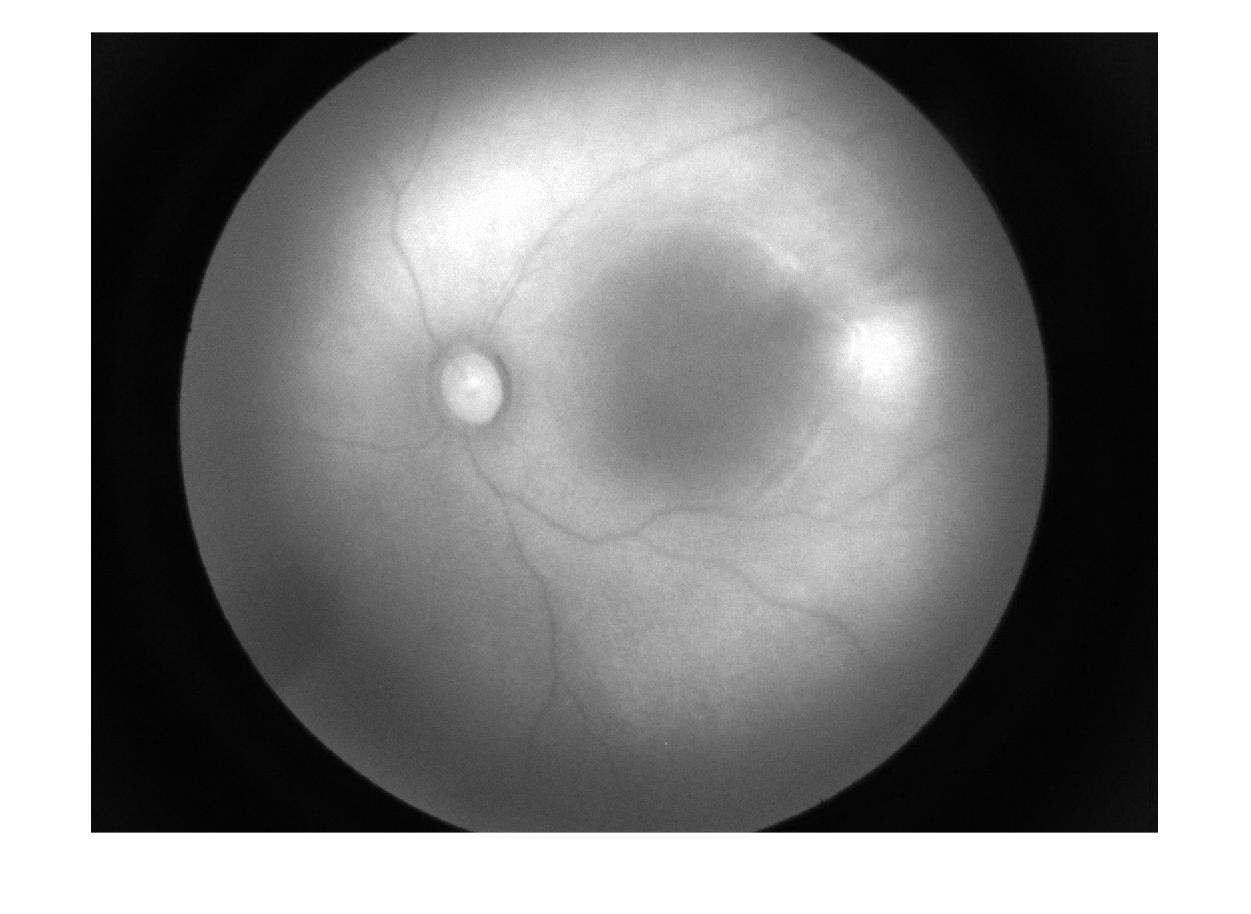}
        \caption{R-channel}
        \label{fig:image features R channel}
	\end{subfigure}%
	\hfill
	\begin{subfigure}{.34\linewidth}
		\centering
        \includegraphics[width=1\linewidth,height=4cm]{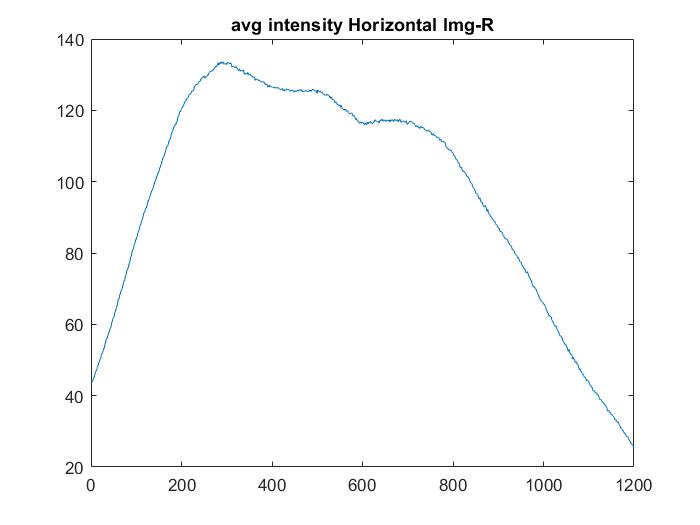}
        \caption{Horizontal avg.}
        \label{fig: horizontal Avg. of R channel}
	\end{subfigure}	
	\hfill
	\begin{subfigure}{.34\linewidth}
		\centering
        \includegraphics[width=1\linewidth,height=4cm]{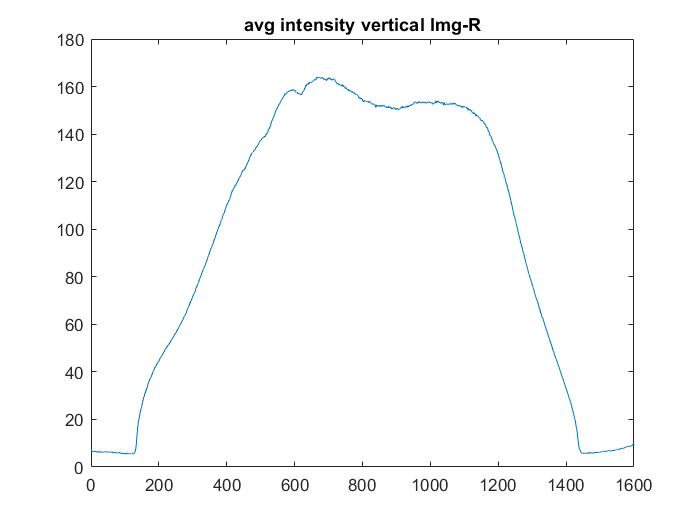}
        \caption{Vertical avg.}
        \label{fig:vertical Avg. of R channel}
	\end{subfigure}
    \hfill
    \begin{subfigure}{0.30\linewidth}
		\centering
        \includegraphics[width=1\linewidth, height=4cm]{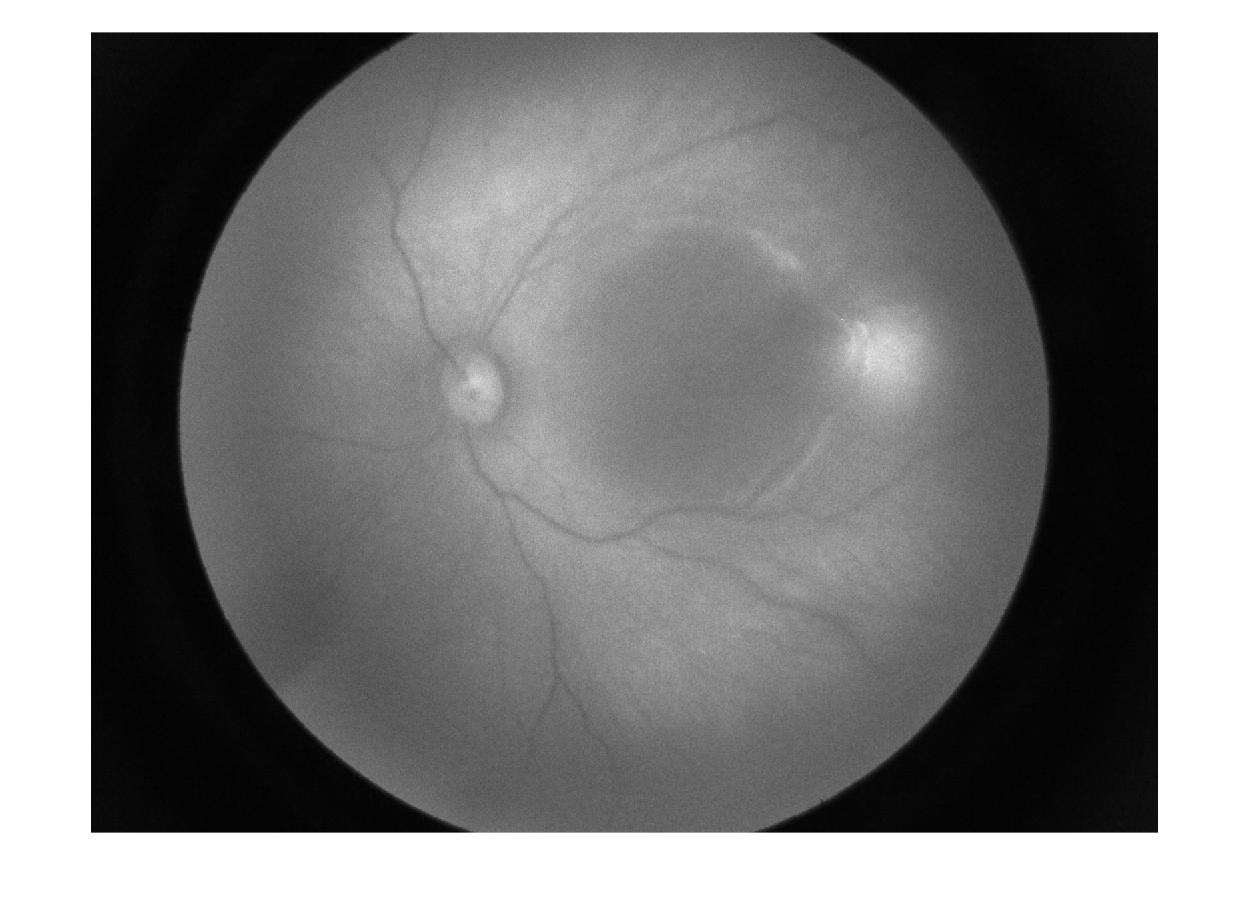}
        \caption{G-channel}
        \label{fig:image features G channel}
	\end{subfigure}%
	\hfill
	\begin{subfigure}{.34\linewidth}
		\centering
        \includegraphics[width=1\linewidth,height=4cm]{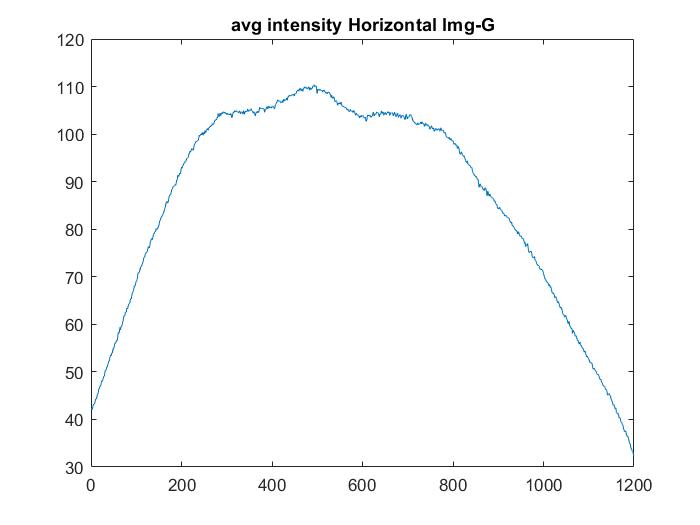}
        \caption{Horizontal avg.}
        \label{fig: horizontal Avg. of G channel}
	\end{subfigure}	
	\hfill
	\begin{subfigure}{.34\linewidth}
		\centering
        \includegraphics[width=1\linewidth,height=4cm]{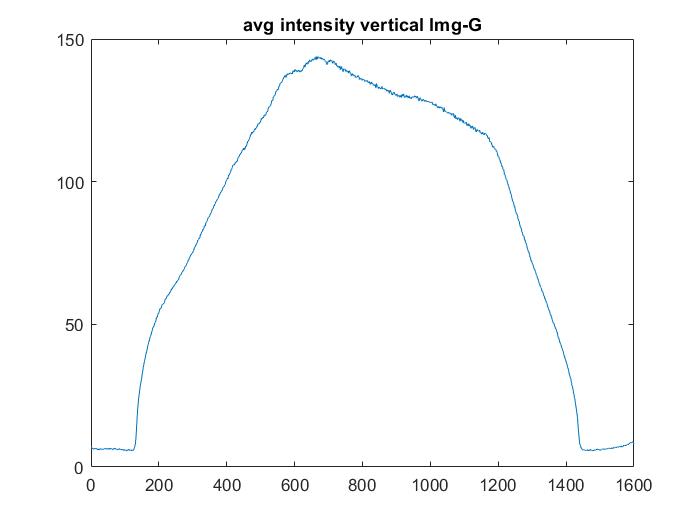}
        \caption{Vertical avg.}
        \label{fig:vertical Avg. of G channel}
	\end{subfigure}
    \hfill
    \begin{subfigure}{0.3\linewidth}
		\centering
        \includegraphics[width=1\linewidth, height=4cm]{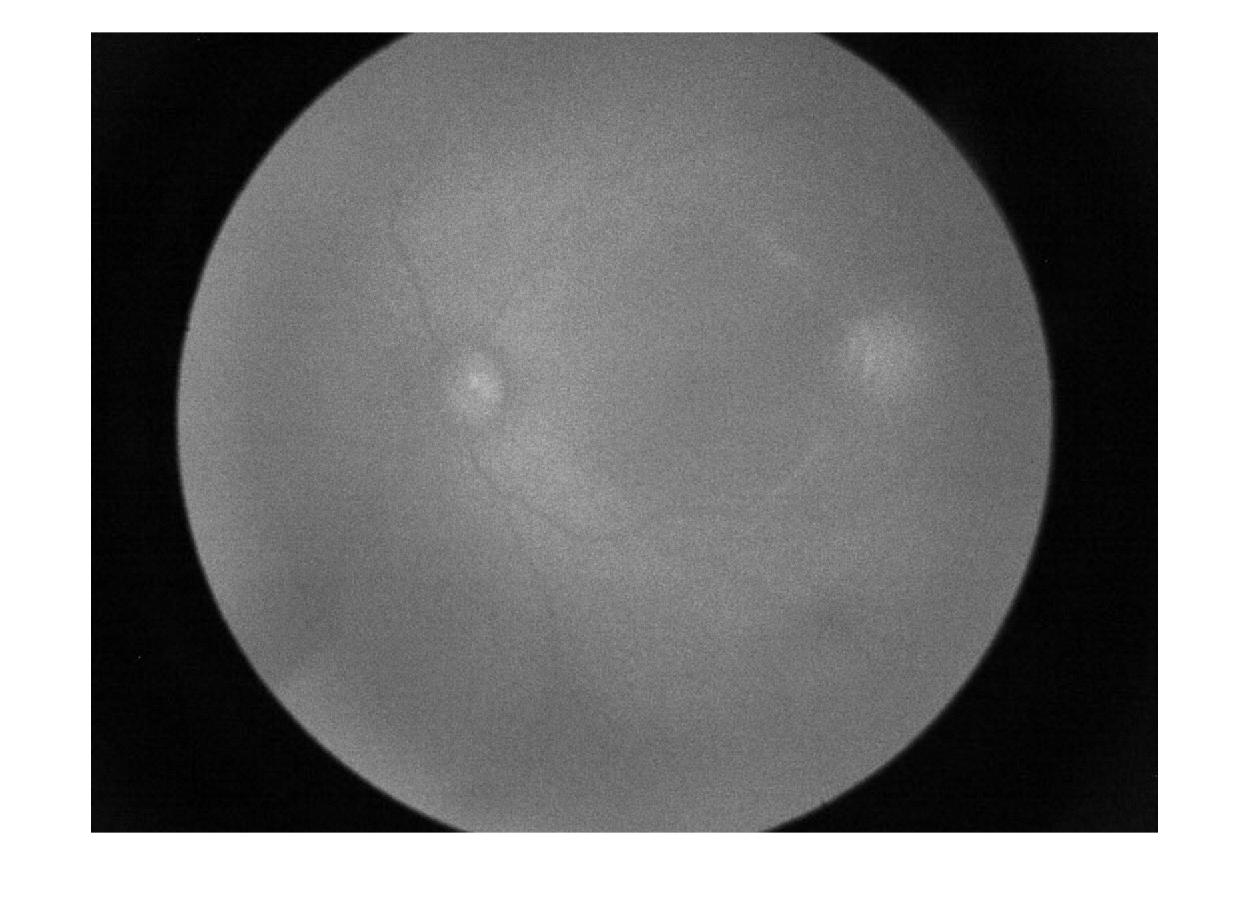}
        \caption{B-channel}
        \label{fig:image features B channel}
	\end{subfigure}%
	\hfill
	\begin{subfigure}{.34\linewidth}
		\centering
        \includegraphics[width=1\linewidth,height=4cm]{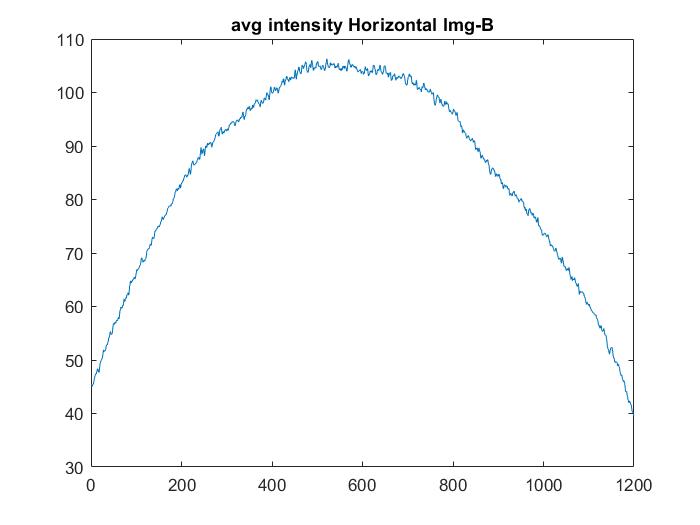}
        \caption{Horizontal avg.}
        \label{fig: horizontal Avg. of B channel}
	\end{subfigure}	
	\hfill
	\begin{subfigure}{.34\linewidth}
		\centering
        \includegraphics[width=1\linewidth,height=4cm]{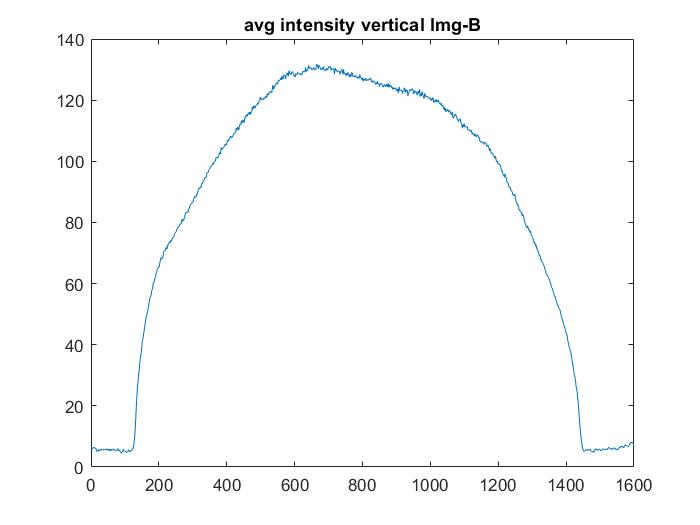}
        \caption{Vertical avg.}
        \label{fig:vertical Avg. of B channel}
	\end{subfigure}
 
	\caption{Image feature extraction by averaging horizontal and vertical components and plotting their spatial extent for three different channels of an image. All channels have a unique spatial average intensity variation represented in the plot, enabling reliable identification and distinction between images and color channels.}
	\label{fig:fundus image feature for DFP}
\end{figure}
\begin{table}[t!]
\centering
\caption{ Correlation between different fundus image features captured by an imaging device for a person's eye at different instances.}
\label{tab: Correlation between the different instants for a same person}
\begin{tabular}{|l|l|l|l|l|l|l|l|l|l|l|}
\hline
 & Img-1 & Img-2 & Img-3 & Img-4 & Img-5 & Img-6 & Img-7 & Img-8 & Img-9 & Img-10 \\ \hline
Img-1 & \textbf{1.000} & 0.850 & 0.778 & 0.899 & 0.959 & 0.975 & 0.874 & 0.838 & 0.838 & 0.794 \\ \hline
Img-2 & 0.850 &\textbf{ 1.000} & 0.946 & 0.985 & 0.928 & 0.907 & 0.989 & 0.980 & 0.990 & 0.984 \\ \hline
Img-3 & 0.778 & 0.946 & \textbf{1.000} & 0.925 & 0.847 & 0.826 & 0.932 & 0.980 & 0.976 & 0.978 \\ \hline
Img-4 & 0.899 & 0.985 & 0.925 &\textbf{ 1.000} & 0.962 & 0.946 & 0.987 & 0.964 & 0.975 & 0.953 \\ \hline
Img-5 & 0.959 & 0.928 & 0.847 & 0.962 & \textbf{1.000} & 0.995 & 0.952 & 0.903 & 0.910 & 0.876 \\ \hline
Img-6 & 0.975 & 0.907 & 0.826 & 0.946 & 0.995 & \textbf{1.000} & 0.930 & 0.885 & 0.889 & 0.852 \\ \hline
Img-7 & 0.874 & 0.989 & 0.932 & 0.987 & 0.952 & 0.930 & \textbf{1.000} & 0.972 & 0.980 & 0.966 \\ \hline
Img-8 & 0.838 & 0.980 & 0.980 & 0.964 & 0.903 & 0.885 & 0.972 & \textbf{1.000} & 0.995 & 0.992 \\ \hline
Img-9 & 0.838 & 0.990 & 0.976 & 0.975 & 0.910 & 0.889 & 0.980 & 0.995 & \textbf{1.000} & 0.995 \\ \hline
Img-10 & 0.794 & 0.984 & 0.978 & 0.953 & 0.876 & 0.852 & 0.966 & 0.992 & 0.995 & \textbf{1.000} \\ \hline
\end{tabular}%
\end{table}
\subsubsection{Fingerprinting of an Image}
The inherent randomness of the imaging sensor, hardware, software, and imaging conditions have been utilized to generate a unique identifier for each image in a medical dataset. We specifically take advantage of the variations in sensor pixel responses that cause spatial intensity variation. This is accomplished by computing the horizontal and vertical average intensity profiles of an image, which produce a unique signature (see Figure ~\ref{fig:fundus image feature for DFP} for example, the image of fundus). Table ~\ref{tab: Correlation between the different instants for a same person} shows the watermarked embedded features of an image with the images having maximum similarity that are acquired for the same person using a device at a different instant of time and show the correlation coefficient is equal to \textbf{1.0} for the same image.

\begin{table}[b!]
\centering
\caption{Comparative analysis of image quality in the proposed framework and diverse medical image datasets, showcasing the superior performance of the proposed approach.
}
\resizebox{1\linewidth}{!}{%
\begin{tabular}{ccccccccc}\hline
\multicolumn{1}{|c|}{\multirow{2}{*}{Evaluation matrix}} & \multicolumn{8}{c|}{Image dataset} \\ \cline{2-9} 
\multicolumn{1}{|c|}{} & \multicolumn{1}{c|}{ROP   Dataset}  & \multicolumn{1}{c|}{Blood   cell}  & \multicolumn{1}{c|}{Dermne   Acne}  & \multicolumn{1}{c|}{Dermne   hair loss}  & \multicolumn{1}{c|}{Dermne-nail-fungus}  & \multicolumn{1}{c|}{wound-foot-ulcer}  & \multicolumn{1}{c|}{Skin-lesion}  & \multicolumn{1}{c|}{Dermne-Actinic}  \\ \hline

\multicolumn{1}{|l|}{MSE} & \multicolumn{1}{l|}{11.004$\pm$0.393}  & \multicolumn{1}{l|}{10.674$\pm$0.154}  & \multicolumn{1}{l|}{11.337$\pm$1.122}  & \multicolumn{1}{l|}{10.531$\pm$0.593}  & \multicolumn{1}{l|}{11.411$\pm$0.836}  & \multicolumn{1}{l|}{12.583$\pm$0.659}  & \multicolumn{1}{l|}{13.317$\pm$1.353}  & \multicolumn{1}{l|}{10.549$\pm$0.476}  \\ \hline

\multicolumn{1}{|l|}{RMSE} & \multicolumn{1}{l|}{3.317$\pm$0.059}  & \multicolumn{1}{l|}{3.267$\pm$0.024}  & \multicolumn{1}{l|}{3.363$\pm$0.162} &  \multicolumn{1}{l|}{3.244$\pm$0.090}  & \multicolumn{1}{l|}{3.376$\pm$0.123} & \multicolumn{1}{l|}{3.546$\pm$0.093}  & \multicolumn{1}{l|}{3.645$\pm$0.182}  & \multicolumn{1}{l|}{3.247$\pm$0.072}  \\ \hline

\multicolumn{1}{|l|}{PSNR} & \multicolumn{1}{l|}{37.718$\pm$0.157} & \multicolumn{1}{l|}{37.848$\pm$0.063} & \multicolumn{1}{l|}{37.605$\pm$0.407} & \multicolumn{1}{l|}{37.912$\pm$0.238} & \multicolumn{1}{l|}{37.569$\pm$0.313} & \multicolumn{1}{l|}{37.138$\pm$0.227} & \multicolumn{1}{l|}{36.906$\pm$0.425} & \multicolumn{1}{l|}{37.903$\pm$0.192} \\ \hline
\multicolumn{1}{|l|}{UQI} & \multicolumn{1}{l|}{0.963$\pm$0.005} & \multicolumn{1}{l|}{1.000$\pm$0.000} & \multicolumn{1}{l|}{0.990$\pm$0.018} & \multicolumn{1}{l|}{0.992$\pm$0.020} & \multicolumn{1}{l|}{0.860$\pm$0.135} & \multicolumn{1}{l|}{0.760$\pm$0.002} & \multicolumn{1}{l|}{0.975$\pm$0.078} & \multicolumn{1}{l|}{0.977$\pm$0.052} \\ \hline
\multicolumn{1}{|l|}{SSIM} & \multicolumn{1}{l|}{0.927$\pm$0.003} & \multicolumn{1}{l|}{0.927$\pm$0.002} & \multicolumn{1}{l|}{0.960$\pm$0.015} & \multicolumn{1}{l|}{0.987$\pm$0.012} & \multicolumn{1}{l|}{0.907$\pm$0.046} & \multicolumn{1}{l|}{0.857$\pm$0.008} & \multicolumn{1}{l|}{0.897$\pm$0.029} & \multicolumn{1}{l|}{0.960$\pm$0.015} \\ \hline
\multicolumn{1}{|l|}{ERGAS} & \multicolumn{1}{l|}{7815.419$\pm$710.325} & \multicolumn{1}{l|}{780.167$\pm$10.149} & \multicolumn{1}{l|}{3755.320$\pm$5778.526} & \multicolumn{1}{l|}{4610.039$\pm$9675.071} & \multicolumn{1}{l|}{51596.039$\pm$56128.958} & \multicolumn{1}{l|}{1439.984$\pm$556.671} & \multicolumn{1}{l|}{7269.485$\pm$20359.152} & \multicolumn{1}{l|}{6062.707$\pm$8162.849} \\ \hline
\multicolumn{1}{|l|}{SCC} & \multicolumn{1}{l|}{0.850$\pm$0.013} & \multicolumn{1}{l|}{0.619$\pm$0.008} & \multicolumn{1}{l|}{0.930$\pm$0.041} & \multicolumn{1}{l|}{0.981$\pm$0.026} & \multicolumn{1}{l|}{0.781$\pm$0.126} & \multicolumn{1}{l|}{0.504$\pm$0.059} & \multicolumn{1}{l|}{0.553$\pm$0.075} & \multicolumn{1}{l|}{0.903$\pm$0.072} \\ \hline
\multicolumn{1}{|l|}{SAM} & \multicolumn{1}{l|}{0.028$\pm$0.004} & \multicolumn{1}{l|}{0.017$\pm$0.000} & \multicolumn{1}{l|}{0.026$\pm$0.006} & \multicolumn{1}{l|}{0.033$\pm$0.008} & \multicolumn{1}{l|}{0.026$\pm$0.010} & \multicolumn{1}{l|}{0.029$\pm$0.005} & \multicolumn{1}{l|}{0.022$\pm$0.003} & \multicolumn{1}{l|}{0.028$\pm$0.007} \\ \hline
\multicolumn{1}{|l|}{MSSSIM} & \multicolumn{1}{l|}{0.955$\pm$0.007} & \multicolumn{1}{l|}{0.970$\pm$0.001} & \multicolumn{1}{l|}{0.978$\pm$0.007} & \multicolumn{1}{l|}{0.990$\pm$0.004} & \multicolumn{1}{l|}{0.980$\pm$0.006} & \multicolumn{1}{l|}{0.979$\pm$0.003} & \multicolumn{1}{l|}{0.948$\pm$0.008} & \multicolumn{1}{l|}{0.984$\pm$0.004} \\ \hline
\multicolumn{1}{|l|}{VIFP} & \multicolumn{1}{l|}{0.606$\pm$0.008} & \multicolumn{1}{l|}{0.618$\pm$0.005} & \multicolumn{1}{l|}{0.681$\pm$0.040} & \multicolumn{1}{l|}{0.765$\pm$0.021} & \multicolumn{1}{l|}{0.689$\pm$0.029} & \multicolumn{1}{l|}{0.688$\pm$0.015} & \multicolumn{1}{l|}{0.566$\pm$0.036} & \multicolumn{1}{l|}{0.696$\pm$0.024} \\ \hline
\multicolumn{1}{|l|}{PSNRB} & \multicolumn{1}{l|}{33.809$\pm$0.327} & \multicolumn{1}{l|}{33.786$\pm$0.110} & \multicolumn{1}{l|}{34.550$\pm$0.784} & \multicolumn{1}{l|}{35.969$\pm$0.617} & \multicolumn{1}{l|}{34.278$\pm$1.086} & \multicolumn{1}{l|}{33.751$\pm$0.268} & \multicolumn{1}{l|}{32.668$\pm$0.388} & \multicolumn{1}{l|}{34.828$\pm$0.406} \\ \hline
\end{tabular}%
}
\label{tab: comparison of image quality}
\end{table}

\subsubsection{Feasibility Analysis of Framework}
In this, we first study the quality of watermarked images with payload information as DFP. Table~\ref{tab: comparison of image quality} summarizes the effect on the image quality after image watermarking in the proposed framework for different medical image datasets. We consider various metrics used to evaluate image quality, such as PSNR, SSIM, MSE, RMSE, etc. Results show that the quality of the original image is preserved as we have used a hybrid method of image watermarking. 
\par
\begin{table}[t!]
\centering
\caption{Dataset-wise time analysis for different stages in the proposed medical image data provenance framework.
}
\resizebox{\linewidth}{!}{%
\begin{tabular}{|c|c|c|c|c|c|c|c|c|}
\hline
\multirow{2}{*}{Stage name} & \multicolumn{8}{c|}{Datasets   wise time analysis (in Second)} \\ \cline{2-9}

 &   ROP Dataset & Blood cell & Dermne Acne & Dermne hair loss & Dermne-nail-fungus & wound-foot-ulcer & Skin-lesion & Dermne-Actinic \\ \hline
 
Feature   extraction & {1.4725}  $\pm${0.0005} & {0.0006}  $\pm$ {0.0007} & {0.0009}  $\pm$ {0.0009} & {0.0011}  $\pm$ {0.0014} & {0.0006}  $\pm$ {0.0010} & {0.0004}  $\pm$ {0.0005} & {0.0026}  $\pm$ {0.0032} & {0.0007} $\pm$ 0.0010 \\ \hline

DFP embedding & {6.7687}  $\pm$ {0.5229} & {1.0264}  $\pm$ {0.0401} & {1.0863}  $\pm$ {0.0422} & {1.2287}  $\pm$ {0.1604} & {1.1564}  $\pm$ {0.1599} & {0.8253}  $\pm$ {0.0497} & {2.5150}  $\pm$ {0.2285} & {0.8758} $\pm$ 0.0588 \\ \hline

WM extraction and decode & {4.7212}  $\pm$ {0.4792} & {0.6877}  $\pm$ {0.0332} & {0.7505}  $\pm$ {0.0497} & {0.8055}  $\pm$ {0.0797} & {0.8119}  $\pm$ {0.0897} & {0.5870}  $\pm$ {0.0504} & {1.7424}  $\pm$ {0.1977} & {0.6100} $\pm$ 0.0389 \\ \hline

\end{tabular}%
}
\label{tab: Dataset-wise time analysis}
\end{table}
Our study compared the computational overhead for DFP generation, embedding, and extraction. The framework required only a few milliseconds to generate DFPs for most datasets. However, DFP generation took longer than 1 second in the ROP dataset. This was primarily due to the ROP dataset's higher image resolution than the other datasets. For embedding DFPs and extracting or decoding the embedded DFP information from watermarked images, the framework required approximately 1 second. Again, the higher image resolution in the ROP dataset caused the framework to need more time for DFP embedding and extraction/decoding. These analyses do not include setup time for telecommunications or read/write operations for images. Table ~\ref{tab: Dataset-wise time analysis} provides the dataset-wise time analysis for the execution of different stages in the proposed framework.

\subsubsection{Immutability Analysis: Attack and Tamper Detection}~\label{subsub}~\label{subsub}
The immutability of data/images means that they cannot be altered after being created and captured by the healthcare system. Imaging devices are potential attack sites for malicious users or attackers. Therefore, to ensure the immutability of data/images in the healthcare system, we propose a provenance method that analyzes the immutability of data/images under the considered attack model scenario. The proposed method specifically focuses on imaging devices and considers several common types of image processing attacks, including (a) copying and pasting an image, (b) extracting a watermark from the original image and embedding it into a fake image, and (c) cropping a section of a legitimate image and attaching it to a fake image.
Our framework can easily detect a fake image for the first type of attack, where an image is copy-pasted since it lacks unique watermarked information based on the DFP. However, in the other two attack scenarios (extracting watermarks and cropping-based attacks), the information obtained only provides the device source information and fails to validate the uniqueness of the images based on their intrinsic characteristics.
\par 

\begin{figure}[b!]
	\centering
	\includegraphics[width=0.75\linewidth]{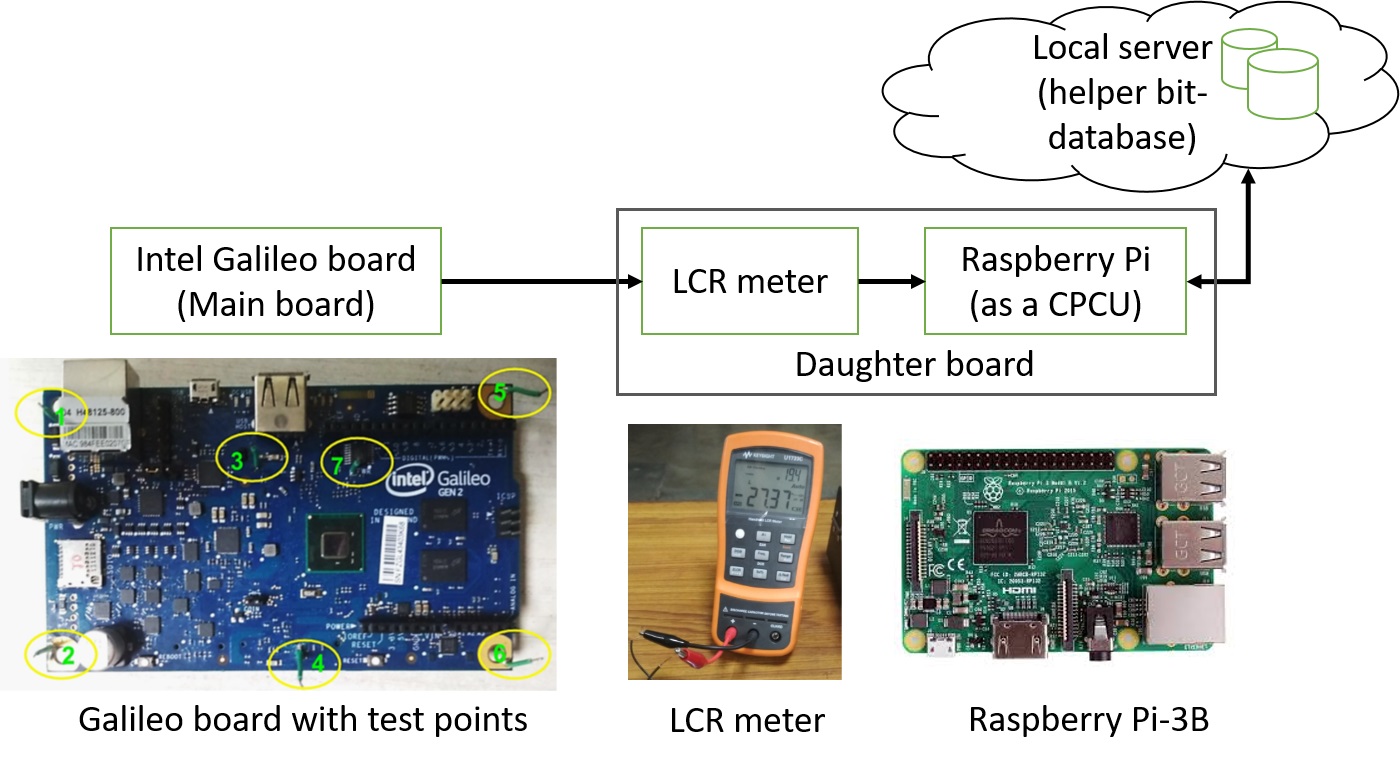}
	\caption{Realisation of the proposed system using Intel Galileo, Raspberry Pi-3B, LCR meter and local cloud server.} 
	\label{fig: realisation of a system}
\end{figure}
\subsection{Hardware-based system for device fingerprinting}~\label{subsubsec: results hardware-based device fingerprint generation}
For the evaluation, we used the electronic board setup shown in Figure ~\ref{fig: realisation of a system}.


\paragraph*{Experimental Setup}
For the measurement of the device characteristic parameter, we have used an LCR meter (Model: GW-instek LCR-916) meter, Raspberry-Pi and local server shown in Figure ~\ref{fig: realisation of a system}, which save measured value logs in the system. The LCR meter and Raspberry Pi board are used to realise the proposed system \textit{DevFing}, and the Galileo board is used as a test board.  We took some initial sets of measurement of LCRZ for the single combination (i.e., 6-set) of test points on the board. Therefore, the electronic board's feature vector $D_{LCR}$ is:
\small{
	\setlength{\arraycolsep}{1.50pt} 
	\medmuskip = 1mu 
	$$\left[
	\begin{matrix}
	253.9e^{-3}&703.9e^{-6}&2.828e^{-3}&2.195e^{-3}&706.6e^{-6}&703.5e^{-6}\\
	997.2e^{-12}&358.7e^{-9}&88.92e^{-9}&114.4e^{-9}&355.0e^{-9}&358.0e^{-9}\\
	9.352e^{3}&0.546e^{3}&1.381e^{3}&1.174e^{3}&0.547e^{3}&0.546e^{3}\\
	1e^{-6}&0.304e^{3}&32.69e^{3}&15.96e^{3}&0.305e^{3}&0.305e^{3}
	\end{matrix}
	\right]
	$$
}
\normalsize
and the final binary stream $D_{binary}$ is:
\small
\begin{equation}
D_{binary}=
\left[
\begin{matrix}
L_{binary}\\
C_{binary}\\
R_{binary}\\
Z_{binary}
\end{matrix}
\right]^{T} =
\left[
\begin{matrix}
1&0&1&0\\
0&1&0&1\\
0&1&1&1\\
0&1&1&1\\
0&1&0&1\\
0&1&0&1
\end{matrix}
\right]=
\left[
\begin{matrix}
0\times A \\
0\times5\\
0\times7\\
0\times7\\
0\times5\\
0\times5
\end{matrix}
\right]
\end{equation}
\normalsize
Further, we used these sets of board features' value to simulate and  tested for the stability of the device signature in the proposed system. For the fuzzy extractor, we have used Python's based Fuzzy Extractor function. In the simulation, we consider the constant value of all parameters  such as  reproduce error: $0.001$ and digital locker arguments like a hash function: $SHA-256$, security parameter and length of bits of nonce used in digital locker: $16$) of the extractor function except the source values, key-length and hamming distance~\cite{dodis2004fuzzy,python2019FuzzyExtractor}.   
\par
For the above analysis of setup-based device fingerprinting techniques, all experimental simulations are performed on an IoT application-grade electronics board (i.e., Raspberry Pi-3B). The accuracy and performance of our methods are discussed in the coming section.

\paragraph*{Performance Analysis}
For the proposed system, the server-based authentication module verifies whether the device belongs to that device, which claims to be of the IoT network, or not. Therefore, whenever a new device is trying to connect with the existing network, the system first classifies that device using their fingerprint as genuine or fake.  Hence, to understand the system's performance, we define some performance metrics. For example, for a system, true-positive authentication is $TP_{n}$, true-negative authentication is $TN_{n}$, false-positive authentication rate is $FP_{n}$ and false-negative authentication is $FN_{n}$. To measure the error-correcting module's performance, we define accuracy as equal to:
\begin{equation}
    Accuracy = \frac{TP_{n}+TN_{n}}{TP_{n}+TN_{n}+FP_{n}+FN_{n}} 
\end{equation}\label{equ:accuracy}
In addition, for our setup, we consider the $TN_{n}$ and $FN_{n}$ values are as zeros.
\par

\begin{figure}[t!]
    \centering
    \begin{minipage}[c]{0.48\linewidth}
        \centering
        \includegraphics[width=1\textwidth]{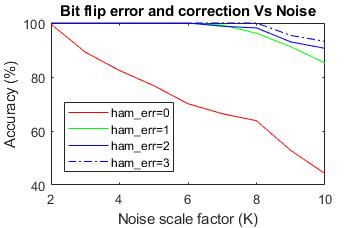}
        \caption{Device signature error (bit flip) corrections accuracy with respect to the noise scale factor at hamming distance equal to $ham\_err$=0 (or without fuzzy extractor) to $ham\_err$=3. 
	}
	\label{fig:bit flip error mitigation}
    \end{minipage}
    \hfill
    \begin{minipage}[c]{0.48\linewidth}
        \centering
	\includegraphics[width=01\textwidth]{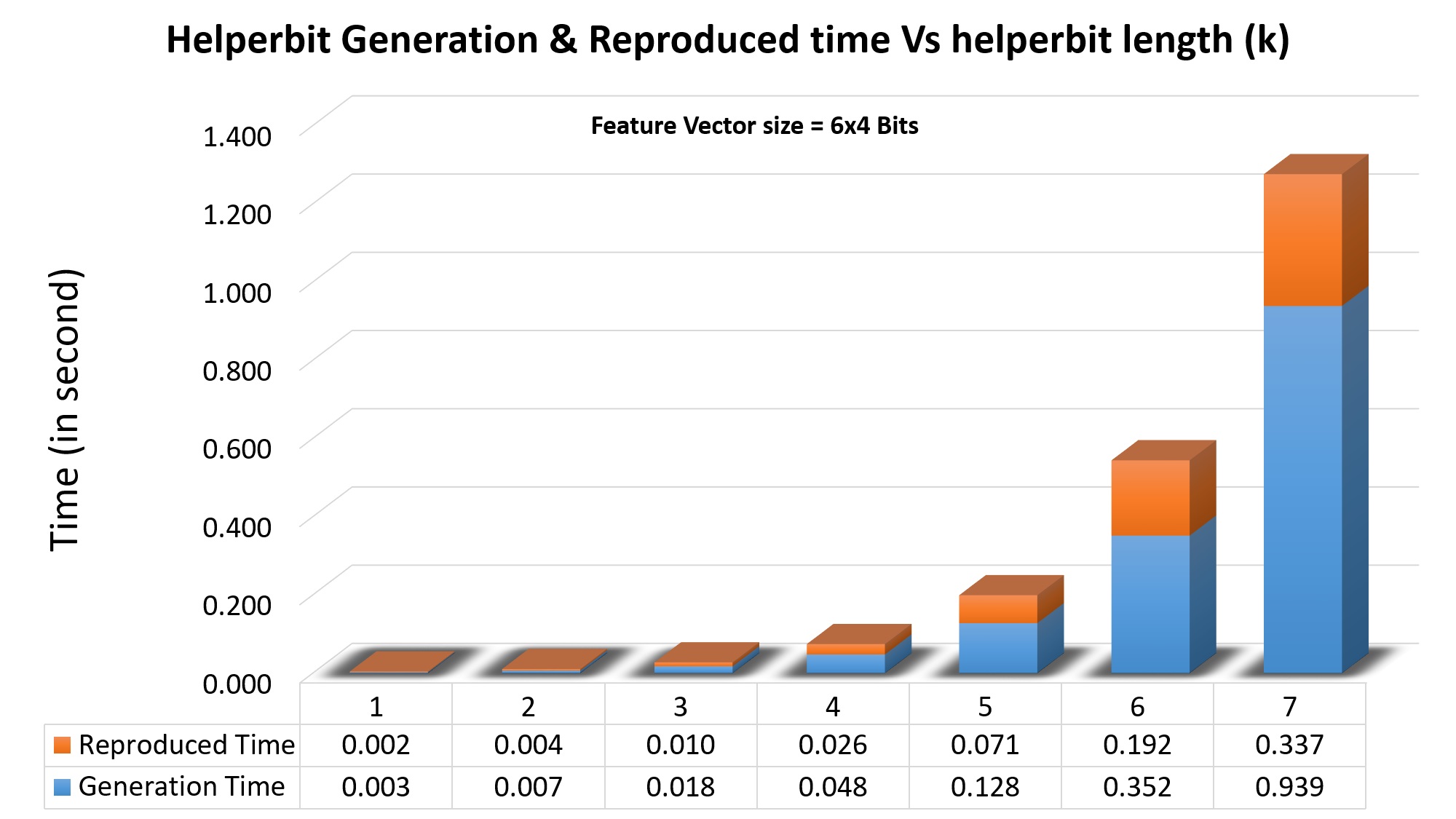}
	\caption{Device signature error correction and overhead analysis with respect to hamming distance ($ham\_err$ = 1 to 7):  time taken by the fuzzy extractor in generation and producer of helper bit for a feature vector of size $6\times4$ bits. 
	}
	\label{fig:overhead and performance analysis with heleper bits size}
    \end{minipage}
\end{figure}
Next, we evaluate the impact of bit-flip errors, the most common error generated due to variations of device-ageing. Here, we used the python code to simulate noise added with the characteristics parameters of the board. It adds the simulated error/noise ($K*\eta$) in log values in the device feature vector ($log(D_{LCR})$), where $K$ is the noise scaling factor to control the amount of noise during the simulation and $\eta$ is the pseudo-random values generator between 0 to 1. As a result, the device's signature bit-stream is also affected by the noisy feature and appears as a bit-flip error. 
\par
This error has been taken care of by a fuzzy extractor-based error correction unit in the system.  In Figure ~\ref{fig:bit flip error mitigation}, we show how the proposed system improves the signature matching accuracy that suffers from bit-flip problems. It achieves accuracy up to 100\% for noise scale (K=8) and hamming distance ($ham\_err$=3).   However, the system has also some overhead that is shown in Figure ~\ref{fig:overhead and performance analysis with heleper bits size}. 

\subsection{Discussion}
\paragraph*{Selection of DFP Methods}
In this study, we discuss two different methods for device fingerprinting-based in the context of data provenance. The first method utilizes the device's electrical characteristics, which is beneficial in systems where new devices are added remotely within a complex interconnected network. This is particularly relevant in medical systems where the importance of these methodologies is high, as counterfeit healthcare devices can lead to breaches of personal health-related information and, in some cases, even jeopardize lives. This system proves to be useful in remote areas with limited resources, where operators may have limited knowledge and physical monitoring is not always feasible.
\par
On the other hand, when it comes to imaging-based healthcare devices, operators capture images that display visible pathological manifestations and share these images with experts for consultation, screening, and disease diagnosis. Particularly in telemedicine/mHealth/IoMT or any connected healthcare setups, the devices used by remote social workers are duly registered within the healthcare system. However, the risk of image tampering and the dissemination of counterfeit images remains substantial in such scenarios. The proposed system is valuable for protecting and preserving trust and source integrity.
\par
Additionally, designing an additional daughterboard for mobile devices, particularly smartphones, is often impractical due to the complexity of board development and the selection of test points. Consequently, in such cases, an application-based approach proves to be a more feasible solution. This approach can capture hardware and software-related information from the device using available APIs and applications, ensuring the robustness of the device fingerprinting process.
\paragraph*{Selection of DFP Features}
The selection of device unique features for generating device fingerprints presents a complex challenge as it requires identifying features that are unique to each device and cannot be replicated or spoofed by malicious users. Feature selection depends on various factors, including the type of device, its intended application, and the specific deployment scenarios. For instance, in the case of mobile medical devices, the choice of features depends on the application and the device's operating environment. A medical imaging-based application that utilizes a camera and light measurement-related sensors to maintain the quality of captured images would require features related to these hardware sensors, peripheral components, as well as firmware and software information. This comprehensive feature selection approach ensures that the device can be accurately and uniquely distinguished for identification purposes.
\par 
However, the challenge of selecting unique features is complicated by the fact that malicious users can attempt to replicate or spoof the selected features to evade detection. Therefore, it is essential to evaluate the robustness of the selected features against various attacks and to incorporate mechanisms to detect and prevent spoofing attempts. The suitability of the chosen features depends on the specific requirements and constraints of the application and the environment in which it operates, as well as the effectiveness of the selected features in preventing spoofing and ensuring device identification.
\paragraph*{Selection of Image Watermarking Method}
In this study, we employed an invisible watermarking technique to minimize its impact on the visibility of processed images. In contrast, visible watermarking techniques directly affect image quality. This effect is particularly pronounced in medical imaging, influencing disease screening and diagnosis accuracy. The ongoing advancements in image processing, computer vision, and machine/deep learning have improved watermarking methods. Consequently, developers and users now have various options to choose the most suitable method, considering factors such as transparency, security, and impact on image quality. This decision is crucial for robust content protection without compromising visual integrity. 

\paragraph*{Comparison with other existing provenance frameworks}
The traditional model for expressing provenance utilizes a graph structure composed of annotated nodes and edges~\cite{glavic2021data,pan2023data,wittner2022lightweight}. In this context, nodes typically denote documented procedures or objects, and edges symbolize their interconnections.
Although some recent works explore blockchain-based methods~\cite{d2022data,de2022blockchain}, these approaches are impractical in low-resource settings. Moreover, the increased use of AI, and ML/DL techniques for data provenance requires significant computational infrastructure, limiting their real-time implementation in low-resource remote healthcare settings~\cite{ahmed2023data,wang2022development,glavic2021data}. Thus, their importance in situations where device users are compromised and with legacy devices may be limited, stressing the importance of our suggested framework in ensuring safe and reliable remote healthcare services in settings with limited resources.

\section{Conclusion}\label{sec:conclusion}
This work explores the requirement for stable, secure, and tamper-proof data provenance and device identification methods in the rapidly evolving landscape of connected devices, particularly in the medical and healthcare sectors. The proposed DFP-based device identification system leverages physical intrinsic features to promptly detect malicious tampering, especially in low-resource settings where the probability of physical tampering of devices and data is high. Additionally, the paper presents a framework for tracking the provenance of medical images, integrating techniques such as device fingerprinting and digital watermarking to embed origin information directly into the images.
This framework is particularly useful in ophthalmic imaging as well as other medical imaging applications in remote and resource-limited areas, ensuring the integrity and authenticity of subsequent tasks such as disease diagnosis, screening, and monitoring through the use of these images. The integration of the original image's spatial characteristics information enhances the framework's resilience against common image tampering attacks. Furthermore, this framework is also tested with other healthcare datasets, considering the potential application of handheld, portable, or wearable imaging devices for healthcare use in imaging-based healthcare services.

\bibliographystyle{unsrt}






\end{document}